\documentclass[12pt,letterpaper]{article}
\usepackage{amsmath,amssymb,array,calc,rotating,epsfig,psfrag, amscd, simplewick}
\usepackage[left=2cm,top=1cm,right=3cm,nohead]{geometry}

\newcommand{\nc}{\newcommand}

\nc{\ra}{\rightarrow} 
\nc{\lra}{\leftrightarrow} 
\nc{\Ra}{\Rightarrow} 
\nc{\LRa}{\Leftightarrow} 
\nc{\blp}{{\big (}}
\nc{\brp}{{\big )}}
\nc{\Blp}{{\Big (}}
\nc{\Brp}{{\Big )}}
\nc{\bglp}{{\bigg (}}
\nc{\bgrp}{{\bigg )}}
\nc{\Bglp}{{\Bigg (}}
\nc{\Bgrp}{{\Bigg )}}
\nc{\slb}{{\rm [}}
\nc{\srb}{{\rm ]}}
\nc{\bslb}{{\rm \big [}}
\nc{\bsrb}{{\rm \big ]}}
\nc{\Bslb}{{\rm \Big [}}
\nc{\Bsrb}{{\rm \Big ]}}
\def\al{\alpha}

\def\eps{\epsilon}
\nc{\veps}{\varepsilon}
\def\gam{\gamma}

\def\lam{\lambda}
\def\om{\omega}
\nc{\vphi}{\varphi}
\def\tha{\theta}
\def\sig{\sigma}

\def\Gam{\Gamma}

\def\Om{\Omega}
\def\Sig{\Sigma}

\nc{\bea}{\begin{eqnarray}}
\nc{\eea}{\end{eqnarray}}
\nc{\be}{\begin{equation}}
\nc{\ee}{\end{equation}}
\nc{\barr}{\begin{array}}
\nc{\earr}{\end{array}}
\nc{\cA}{{\cal A}}
\nc{\cB}{ \cal B}

\def\cD{{\cal D}}

\nc{\cF}{{\cal F}}
\nc{\cG}{{\cal G}}

\nc{\cL}{{\cal L}}
\nc{\cM}{{\cal M}}
\def\N{{\cal N}}

\nc{\cQ}{{\cal Q}}
\nc{\cR}{{\cal R}}

\def\cV{{\cal V}}
\def\cV{{\cal V}}

\def\cZ{{\cal Z}}
\nc{\cQd}{\cQ^{\dagger}}
\nc{\cRd}{\cR^{\dagger}}
\nc{\BB}{{\mathbb B}}
\nc{\CC}{{\mathbb C}}
\nc{\DD}{{\mathbb D}}
\nc{\EE}{{\mathbb E}}
\nc{\FF}{{\mathbb F}}
\nc{\GG}{{\mathbb G}}
\nc{\HH}{{\mathbb H}}
\nc{\JJ}{{\mathbb J}}
\nc{\RR}{{\mathbb R}}
\nc{\PP}{{\mathbb P}}
\nc{\QQ}{{\mathbb Q}}
\nc{\ZZ}{{\mathbb Z}}
\nc{\calone}{{\mathbb 1}}

\nc{\half}{\frac{1}{2}}
\nc{\qrt}{\frac{1}{4}}
\nc{\del}{\partial}

\nc{\delbar}{\bar\partial}
\nc{\thalf}{\frac{t}{2}}
\nc{\Spin}{\operatorname{Spin}}
\nc{\SO}{\operatorname{SO}}

\nc{\Sp}{{\rm Sp}}
\nc{\com}[2]{{ \left[ #1, #2 \right] }}
\nc{\acom}[2]{{ \left\{ #1, #2 \right\} }}
\nc{\rr}{\rightarrow}
\nc{\p}{\partial}
\nc{\LT}{{\LL_\T}}
\nc{\Tr}{{\rm Tr}}
\nc{\tr}{{\rm tr}}
\def\com#1#2{{ \left[ #1, #2 \right] }}
\def\acom#1#2{{ \left\{ #1, #2 \right\} }}
\nc{\Adag}{A^{\dagger}}
\nc{\AdagI}{A^{\dagger I}}
\nc{\AdagJ}{A^{\dagger J}}
\nc{\AdagK}{A^{\dagger K}}
\nc{\AdagL}{A^{\dagger L}}
\nc{\AdagM}{A^{\dagger M}}
\nc{\Bdag}{B^{\dagger}}
\nc{\BdagI}{B^{\dagger}_I}
\nc{\BdagJ}{B^{\dagger}_J}
\nc{\BdagK}{B^{\dagger}_K}
\nc{\BdagL}{B^{\dagger}_L}
\nc{\BdagM}{B^{\dagger}_M}
\nc{\Cdag}{C^{\dagger}}
\nc{\CdagI}{C^{\dagger I}}
\nc{\CdagJ}{C^{\dagger J}}
\nc{\CdagK}{C^{\dagger K}}
\nc{\Ddag}{D^{\dagger}}
\nc{\DdagI}{D^{\dagger I}}
\nc{\DdagJ}{D^{\dagger J}}
\nc{\DdagK}{D^{\dagger K}}
\nc{\ttha}{\tilde{\theta}}
\nc{\tphi}{\tilde{\phi}}
\nc{\tsig}{\tilde{\sig}}
\nc{\tom}{\tilde{\om}}
\nc{\tlam}{\tilde{\lam}}
\nc{\tSig}{\widetilde{\Sig}}
\nc{\tPhi}{\tilde{\Phi}}
\nc{\tPhibar}{\ol{\tPhi}}
\nc{\tPi}{\tilde{\Pi}}
\nc{\tpsi}{\tilde{\psi}}
\nc{\tPsi}{\tilde{\Psi}}
\nc{\tgam}{\tilde{\gam}}
\nc{\tGam}{\tilde{\Gam}}
\nc{\Xit}{\tilde{\Xi}}
\nc{\tb}{\tilde b}
\nc{\tc}{\tilde c}
\nc{\te}{\tilde e}
\nc{\tf}{\tilde f}
\nc{\tg}{\tilde g}
\nc{\tj}{\tilde j}
\nc{\tp}{\widetilde{p}}
\nc{\tq}{\widetilde{q}}
\nc{\ts}{{\tilde s}}
\nc{\tu}{{\tilde u}}
\nc{\tv}{{\tilde v}}
\nc{\tw}{{\tilde w}}
\nc{\tx}{{\tilde x}}
\nc{\ty}{{\tilde y}}
\nc{\tz}{\tilde z}
\nc{\tA}{{\tilde A}}
\nc{\tAbar}{{\ol \tA}}
\nc{\tD}{{\tilde D}}
\nc{\tE}{{\tilde E}}
\nc{\tG}{{\tilde G}}
\nc{\tH}{{\tilde H}}
\nc{\tJ}{{\tilde J}}
\nc{\tJbar}{{\ol {\tilde J}}}
\nc{\tM}{{\tilde M}}
\nc{\tN}{{\tilde N}}
\nc{\tP}{{\tilde P}}
\nc{\tQ}{{\tilde Q}}
\nc{\tR}{{\tilde R}}
\nc{\tS}{\tilde{S}}
\nc{\tF}{\tilde{{\cal F}}}
\nc{\tX}{\widetilde{X}}
\nc{\tcZ}{\tilde{\cZ}}
\nc{\tcZbar}{\ol{\tcZ}}
\nc{\hb}{\hat b}
\nc{\hc}{\hat c}
\nc{\hd}{\hat d}
\nc{\he}{\hat e}
\nc{\hf}{\hat f}
\nc{\hg}{\hat g}
\nc{\hh}{\hat h}
\nc{\hp}{\hat p}
\nc{\hs}{\hat s}
\nc{\hv}{\hat v}
\nc{\hw}{\hat w}
\nc{\hx}{\hat x}
\nc{\hy}{\hat y}
\nc{\hz}{\hat z}
\nc{\zhat}{\hat z}
\nc{\hA}{\widehat{A}}
\nc{\hE}{\widehat{E}}
\nc{\hF}{\widehat{F}}
\nc{\hH}{\widehat{H}}
\nc{\hJ}{\widehat{J}}
\nc{\tK}{\widetilde{K}}
\nc{\hM}{\widehat M}
\nc{\hV}{\widehat V}
\nc{\hcV}{\widehat \cV}

\nc{\ha}{\widehat \alpha}
\nc{\hphi}{\hat{\phi}}
\nc{\hpsi}{\hat{\psi}}
\nc{\hgam}{\hat{\gam}}
\nc{\hPhi}{\hat{\Phi}}
\nc{\hPsi}{\hat{\Psi}}
\nc{\hGam}{\hat{\Gam}}
\nc{\omhat}{\hat{\om}}

\nc{\w}{\wedge}

\nc{\vb}{\vec b}
\nc{\vc}{\vec c}
\nc{\vd}{\vec d}
\nc{\ve}{\vec e}
\nc{\vf}{\vec f}
\nc{\vg}{\vec g}
\nc{\vh}{\vec h}
\nc{\vp}{\vec p}
\nc{\vq}{\vec q}
\nc{\vr}{\vec r}
\nc{\vs}{\vec s}
\nc{\vv}{\vec v}
\nc{\vw}{\vec w}
\nc{\vx}{\vec x}
\nc{\vy}{\vec y}
\nc{\vz}{\vec z}

\nc{\ol}{\overline}
\nc{\abar}{\ol{a}}
\nc{\bbar}{\ol{b}}
\nc{\cbar}{\ol{c}}
\nc{\dbar}{\ol{d}}
\nc{\ebar}{\ol{e}}
\nc{\ibar}{\ol{\imath}}
\nc{\jbar}{\ol{\jmath}}
\nc{\kbar}{\ol{k}}
\nc{\lbar}{\ol{l}}
\nc{\mbar}{\ol{m}}
\nc{\nbar}{\ol{n}}
\nc{\pbar}{\ol{p}}
\nc{\qbar}{\ol{q}}
\nc{\ubar}{\ol{u}}
\nc{\vbar}{\ol{v}}
\nc{\wbar}{\ol{w}}
\nc{\xbar}{\ol{x}}
\nc{\ybar}{\ol{y}}
\nc{\zbar}{\ol{z}}

\nc{\Abar}{\ol{A}}
\nc{\Bbar}{\ol{B}}
\nc{\Cbar}{\ol{C}}
\nc{\Dbar}{\ol{D}}
\nc{\Ebar}{\ol{E}}
\nc{\Fbar}{\ol{F}}
\nc{\Jbar}{\ol{J}}
\nc{\Kbar}{\ol{K}}
\nc{\Lbar}{\ol{L}}
\nc{\Pbar}{\ol{P}}
\nc{\Qbar}{\ol{Q}}
\nc{\Rbar}{\ol{R}}
\nc{\Sbar}{\ol{S}}
\nc{\Tbar}{\ol{T}}
\nc{\Ubar}{\ol{U}}
\nc{\Vbar}{\ol{V}}
\nc{\Wbar}{\ol{W}}
\nc{\Xbar}{{\overline X}}
\nc{\Ybar}{{\overline Y}}
\nc{\Zbar}{{\overline Z}}
\nc{\cZbar}{{\overline \cZ}}

\nc{\epsbar}{\ol{\epsilon}}
\nc{\lambar}{\ol{\lambda}}
\nc{\psibar}{\ol{\psi}}
\nc{\Psibar}{\ol{\Psi}}
\nc{\phibar}{\ol{\phi}}
\nc{\Phibar}{\ol{\Phi}}
\nc{\chibar}{\ol{\chi}}
\nc{\mubar}{\ol{\mu}}
\nc{\nubar}{\ol{\nu}}
\nc{\rhobar}{\ol{\rho}}
\nc{\ombar}{\ol{\om}}
\nc{\Ombar}{\ol{\Om}}
\nc{\Deltabar}{\ol{\Delta}}
\nc{\Thetabar}{\ol{\Theta}}
\nc{\Xibar}{\ol{\Xi}}

\nc{\Dthbar}{\ol{\rm D3}}

\nc{\gdot}{\dot{g}}
\nc{\xdot}{\dot{x}}
\nc{\ydot}{\dot{y}}
\nc{\sinp}{s_{\phi}}
\nc{\cosp}{c_{\phi}}
\nc{\tanp}{t_{\phi}}
\nc{\spone}{s_{\phi_1}}
\nc{\cpone}{c_{\phi_1}}
\nc{\tpone}{t_{\phi_1}}
\nc{\sptwo}{s_{\phi_2}}
\nc{\cptwo}{c_{\phi_2}}
\nc{\tptwo}{t_{\phi_2}}
\nc{\spth}{s_{\phi_3}}
\nc{\cpth}{c_{\phi_3}}
\nc{\tpth}{t_{\phi_3}}
\nc{\calp}{c_{\al}}
\nc{\salp}{s_{\al}}

\nc{\csch}{{\rm csch}}
\nc{\sech}{{\rm sech}}

\nc{\bah}{{\mathbf {\hat{A}}}}
\nc{\bX}{{\mathbf X}}
\nc{\ba}{{\bf a}}
\nc{\bb}{{\bf b}}
\nc{\bc}{{\bf c}}
\nc{\bd}{{\bf d}}
\nc{\bg}{{\bf g}}
\nc{\bk}{{\bf k}}
\nc{\bl}{{\bf l}}
\nc{\bm}{{\bf m}}
\nc{\bn}{{\bf n}}
\nc{\bo}{{\bf o}}
\nc{\bp}{{\bf p}}
\nc{\bq}{{\bf q}}
\nc{\br}{{\bf r}}
\nc{\bs}{{\bf s}}
\nc{\bt}{{\bf t}}
\nc{\bu}{{\bf u}}
\nc{\bv}{{\bf v}}
\nc{\bw}{{\bf w}}
\nc{\bx}{{\bf x}}
\nc{\by}{{\bf y}}
\nc{\bz}{{\bf z}}
\nc{\bom}{{\bf \om}}
\nc{\bombar}{{\mathbf \ombar}}
\nc{\bPhi}{{\bf \Phi}}

\nc{\rma}{{\rm a}}
\nc{\rmb}{{\rm b}}
\nc{\rmc}{{\rm c}}
\nc{\rmd}{{\rm d}}
\nc{\rmg}{{\rm g}}
\nc{\rk}{{\rm k}}
\nc{\rml}{{\rm l}}
\nc{\rmm}{{\rm m}}
\nc{\rmn}{{\rm n}}
\nc{\rmo}{{\rm o}}
\nc{\rmp}{{\rm p}}
\nc{\rmq}{{\rm q}}
\nc{\rmr}{{\rm r}}
\nc{\rms}{{\rm s}}
\nc{\rmt}{{\rm t}}
\nc{\rmu}{{\rm u}}
\nc{\rmv}{{\rm v}}
\nc{\rmw}{{\rm w}}
\nc{\rmx}{{\rm x}}
\nc{\rmy}{{\rm y}}
\nc{\rmz}{{\rm z}}

\nc{\dal}{\dot{\al}}
\nc{\thadot}{\dot{\tha}}
\nc{\thab}{\bar{\theta}}
\nc{\thal}{\theta^{\al}}
\nc{\thdal}{\bar{\theta}^{\dal}}

\nc{\thsigthm}{\tha \sigma^m \thab}
\nc{\thsigthn}{\tha \sigma^n \thab}

\nc{\Dal}{D_{\al}}
\nc{\Ddal}{\bar{D}_{\dal}}
\nc{\CDal}{{\cal D}_{\al}}
\nc{\CDdal}{\bar{\cal D}_{\dal}}

\nc{\eq}[1]{(\ref{#1})}
\nc{\non}{\nonumber}

\nc{\equ}{{\rm eq}}
\def\Im{{\rm Im ~}}
\def\Re{{\rm Re ~}}
\nc{\AdS}{{\rm AdS}}
\nc{\vol}{{\rm vol}}
\nc{\Ainf}{A_{\infty}}
\nc{\End}{{\rm End}}
\nc{\Ext}{{\rm Ext}}
\nc{\IIB}{{\rm IIB}}
\nc{\Ad}{{\rm Ad}}
\nc{\IIA}{{\rm IIA}}
\nc{\Dslash}{\ensuremath \raisebox{0.025cm}{\slash}\hspace{-0.32cm} D}
\nc{\cDslash}{\ensuremath \raisebox{0.025cm}{\slash}\hspace{-0.32cm} \cD}
\nc{\no}{\!:\!\!}
\nc{\ointdz}{\oint\frac{dz}{2\pi i}}
\nc{\ointdzone}{\oint\frac{dz_1}{2\pi i}}
\nc{\ointdztwo}{\oint\frac{dz_2}{2\pi i}}
\nc{\ointdzb}{\oint\frac{d\zbar}{2\pi i}}
\nc{\ointdzbone}{\oint\frac{d\zbar_1}{2\pi i}}
\nc{\ointdzbtwo}{\oint\frac{d\zbar_2}{2\pi i}}
\nc{\dz}{\frac{dz}{2\pi i}}
\nc{\dzb}{\frac{d\zbar}{2\pi i}}
\nc{\bpm}{\begin{pmatrix}}
\nc{\epm}{\end{pmatrix}}
 \nc{\bitem}{\begin{itemize}}
 \nc{\eitem}{\end{itemize}}

\begin{document}
\begin{center}
\vskip 2 cm

{\Large \bf  Missing Mirrors: Type IIA Supergravity \\
\vskip 0.25 cm 
on the Resolved Conifold} \\

\vskip 1.25 cm 
 Nick Halmagyi \\
\vskip 5mm
Institut de Physique Th\' eorique, \\
CEA Saclay, CNRS-URA 2306, \\
91191 Gif sur Yvette, France \\
\vskip 5mm

nicholas.halmagyi@cea.fr.
\end{center}

\begin{abstract}
We consider massive IIA supergravity on the resolved conifold with $SU(2)_L^2\times U(1)_R$ symmetry and $\N=1$ supersymmetry. A one dimensional family of such regular solutions was found by Brandhuber and we propose this to be the mirror to one dimension of the moduli space of IIB solutions on the deformed conifold found by Butti et al. The remaining dimension of the moduli space of Butti et al contains the baryonic branch of Klebanov-Strassler and we propose that the mirror of this is either some stringy resolution of a family of singular solutions found here or must be entirely non-geometric.
\end{abstract}

\section{Introduction}
String theory on conifold singularities has proved to be a immensely rich area of study for a number of years. The resolution of the singularity provided early  insights into the physics of D-branes \cite{Strominger:1995cz} and with the advent of AdS/CFT the explicit Ricci flat metrics found in \cite{Candelas:1989js} proved to be vital in constructing physically interesting examples of gauge/gravity duality \cite{Klebanov:1998hh, Klebanov:2000hb, Maldacena:2000mw}. Studies of the topological string on the conifold have shed light on geometric transitions \cite{Gopakumar:1998ki, Dijkgraaf:2002fc} which were conjectured to be embedded in the full superstring \cite{Vafa:2000wi}. The conifold has also provided a canonical calculable example of hypermultiplet couplings in four dimensions \cite{Ooguri:1996me, RoblesLlana:2007ae}. In this work we will continue this fine tradition of using the conifold as a guiding example for studying certain geometric aspects of string theory.

The solution of Klebanov and Strassler \cite{Klebanov:2000hb} is the prototypical example of a warped 
Calabi-Yau solution to IIB supergravity \cite{Gubser:2000vg, Grana:2001xn}. The metric on the internal manifold is conformal to the Ricci-flat metric on the deformed conifold \cite{Candelas:1989js}, the dilaton is constant and the three form flux is imaginary self-dual. These are particularly interesting flux backgrounds in string theory since due to Yau's theorem and the work \cite{Giddings:2001yu}, one has an existence proof for the supersymmetry equations. 

A very interesting aspect of the Klebanov-Strassler solution is that it belongs to a nontrivial, two-parameter family of solutions \cite{Gubser:2004qj, Butti:2004pk}. One of these parameters corresponds to the supergravity dual of a baryonic vev and on this branch the metric on the internal manifold is no longer Ricci -flat. The solution along this branch is usually described as an $SU(3)$-structure solution, a condition slightly weaker than $SU(3)$-holonomy, which allows for a much more general class of flux. It is not understood whether warped Calabi-Yau backgrounds in general can have unobstructed modes which preserve only the $SU(3)$-structure conditions but due to the solution \cite{Butti:2004pk} we know that at least two such modes exist when the Calabi-Yau manifold is the deformed conifold with the metric of \cite{Candelas:1989js}.

It was observed in \cite{Casero:2006pt} that one dimension of the space of solutions found in \cite{Butti:2004pk} is particularly simple, on this branch only the metric, dilaton and three-form flux $(g,\vphi,F_3)$ are activated, whereas in general along the whole solutions space all fields except the axion have nontrivial profiles. I will refer to this branch as the ``NS-branch" since it is S-dual to an NS solution of the type first studied in \cite{Strominger:1986uh,Hull:1986kz}. Recently it was shown \cite{Maldacena:2009mw} that the full two dimensional space of solutions can be generated by duality from this simpler branch. It was also pointed out in \cite{Maldacena:2009mw} that this branch provides a precise IIB string theory realization of the conifold transition proposed in \cite{Vafa:2000wi}. 

The point of the current work is to examine the mirror IIA version of the exposition in \cite{Maldacena:2009mw}. The geometric transition in IIA on the conifold was first considered in \cite{Acharya:2000gb, Atiyah:2000zz} as an $S^3\ra \tilde{S}^3$ flop when lifted to M-theory, however this picture is somewhat at odds with the IIB picture presented in \cite{Maldacena:2009mw}. We show here that in fact there exists a family of supergravity solutions, originally found by Brandhuber \cite{Brandhuber:2001kq} which we conjecture provides an exact IIA picture mirror to that of \cite{Maldacena:2009mw}.

The main calculation in the current work is an attempt to generalize the one parameter family of solutions found by Brandhuber to a two parameter family which would be a putative mirror to the entire solution space found in \cite{Butti:2004pk}.  What we will find here is that the most general ansatz in type IIA supergravity, including a mass term ($F_0\neq 0$), which respects the same set of symmetries as its IIB counter part, does not admit regular solutions other than those found in \cite{Brandhuber:2001kq}. We do find however a one parameter family of solutions which are all singular at the origin. It would be interesting if these could be thought of as similar in nature to the singular solution of \cite{Klebanov:2000nc} where the singularity is in fact resolved in the IR. In the solutions found in this work, it would have to be stringy effects which come to the rescue in the IR since we exhaust all possible supergravity solutions with the same symmetries.

Mirror symmetry is a symmetry of string theory with $\N=(2,2)$ worldsheet supersymmetry and many aspects of it  are well  understood for Calabi-Yau manifolds without flux. Evidence in support of the $(2,2)$ worldsheet supersymmetry prevailing with the addition of flux was provided in the nice work \cite{Linch:2006ig}. In this work it was shown using the hybrid formalism \cite{Berkovits:1996bf} that to first order in an expansion in flux, around a Calabi-Yau background, a particular worldsheet $\N=(2,2)$ symmetry is in fact unbroken. Using this worldsheet symmetry they were able to derive the linearization of the target space supersymmetry conditions with flux. The key point to this work is that since a warp factor is generated by the flux, the worldsheet CFT  can no longer be split into two decoupled CFT's for the internal and external space. The surviving $(2,2)$ generators are a combination of currents from the internal CFT and the external CFT. However since there is still some version of $(2,2)$ worldsheet supersymmetry, there should also be some version of mirror symmetry available. 

Mirror symmetry for backgrounds with flux has been studied from the point of view of four dimensional effective actions \cite{Gurrieri:2002wz, Grana:2006hr} but there is still a distinct lack of calculable on-shell examples where some of these ideas can be checked. The SYZ \cite{Strominger:1996it} approach is well motivated from the target space physics and it seems promising that this can be generalized to flux backgrounds \cite{Fidanza:2003zi,Tomasiello:2005bp} but this is difficult to check even in Calabi-Yau examples. Attempts have been made \cite{Becker:2004qh} to explicitly perform the T-dualities in a fashion inspired by \cite{Strominger:1996it} but this seems to break the non-Abelian global symmetries of the problem. Essentially, T-dualizing singular $U(1)$ fibrations is hard.

One generic feature of T-duality in the presence of flux appears to be the generation of so-called non-geometric backgrounds \cite{Hellerman:2002ax, Bouwknegt:2004ap,Dabholkar:2005ve}. These poorly understood backgrounds may well be necessary to provide the correct description of the mirror to the full solution space of  \cite{Butti:2004pk} and it is with that in mind that we have given this paper its title.

This paper is organized as follows: In section 2, we review the solution of \cite{Butti:2004pk} and the insight of \cite{Maldacena:2009mw}. In section 3, we discuss the mirror IIA version and the solutions of \cite{Brandhuber:2001kq}. In section 4, we summarize our attempts to generalize the solutions of \cite{Brandhuber:2001kq} to allow for a non-trivial four form flux $G^{(4)}$. In the conclusions we discuss our perspective on how mirror symmetry may be restored by incorporating non-geometric backgrounds.  In the appendix provide all the details of our computations in $d=11$ supergravity as well as massive IIA supergravity.

\section{The Deformed Conifold in Type IIB Supergravity}

The two parameter family of solutions to IIB supergravity on the deformed conifold \cite{Butti:2004pk} which we are interested in all have $M$ units of $D_5$ Page charge
\be
\cQ^{Page}_{D5}=\frac{1}{4\pi^2 \alpha'}\int_{S^3} F_3 = M.
\ee
It was emphasized recently \cite{Casero:2006pt, Maldacena:2009mw} that a particular one dimensional subspace of these solutions is quite simple. This family has non-trivial profiles for only metric, dilaton and RR three-form ($g,\vphi,F_3$) as opposed to the more general solution of \cite{Butti:2004pk} which has all fields activated except for the axion. This simplified solution has the following form
\bea
ds_{10}^2&=&e^{2A}ds_4^2 + \frac{\alpha' M}{4} ds^2_{M_6} \non \\
\Om_{hol}&=&e^{A}\Omega \non \\
&=& e^{-2\phi_0}\cosh \tau(d\tau +i (\sig_3+\Sig_3)) \w\non \\ 
&&\!\!\!\!\!\!\!\!\!\! \Blp  \frac{i}{\cosh \tau}( \sig_1\w \sig_2 -\Sig_1 \w \Sig_2)- \tanh \tau(-\sig_1\w \Sig_1 + \sig_2\w \Sig_2)  + i(-\Sig_1\w \sig_2 + \sig_1 \w \Sig_2)\Brp \non \\
J&=& -\sig_1 \w \sig_2 (\coth \tau(1-\tau \coth \tau) + c) - \Sig_1 \w \Sig_2 (\coth \tau(1-\tau \coth \tau) - c)  \non \\
&&+ (\sig_1 \w \Sig_2 + \Sig_1 \w \sig_2) \frac{1-\tau \coth \tau}{\sinh  \tau} \label{BBsol} \\
e^{2\phi}&=&e^{2\phi_0} \frac{\sinh^2 \tau}{f^{1/2}c'} \non \\
F_3&=&\frac{\alpha' M}{4} \om_3 \non \\
\om_3 &=&(\sig_3+\Sig_3) ( \sig_1\w \sig_2 -\Sig_1 \w \Sig_2) + \frac{\tau}{\sinh\tau}(-\Sig_1\w \sig_2 + \sig_1 \w \Sig_2)\Brp \non \\
&& -d\tau\w  \frac{\tau \coth\tau-1}{\sinh\tau} \tau(-\sig_1\w \Sig_1 + \sig_2\w \Sig_2)  \non
\eea
The functions $(c(\tau),f(\tau))$ satisfy
\be
\begin{array}{rcl}
f'&=&4c\,\sinh^2 \tau \\
c'&=&\frac{1}{f}\Blp c^2 \sinh^2 \tau -(\tau \cosh \tau-\sinh \tau)^2\Brp \\
\end{array} \label{cfeq}
\ee

The two explicitly known solutions to \eq{cfeq} are 
\be
DC:\ \ \begin{array}{rcl}
c^3&=&3\gam^6 (\cosh \tau \sinh \tau-\tau)/2\\
f&=&\gam^{-1/6} c^4  
\end{array} \label{DCsol}
\ee
(in the limit $\gamma\ra \infty$) and 
\be
CV/MN:\ \ 
\begin{array}{rcl}
c&=&\tau \\
f&=&\tau^2 \sinh^2 \tau -(\tau \cosh \tau-\sinh \tau)^2
\end{array} \label{MNsol}
\ee
but it was shown numerically in \cite{ Butti:2004pk} that there is a one parameter family of regular solutions which interpolates between these two points. 

The solution \eq{DCsol} gives the Ricci flat metric on the deformed conifold with a strictly infinite size $S^3$. This solution still has $M$-units of D5 charge and one can think of the infinite size $S^3$ as necessary to dilute the $F_3$ flux and allow for a Ricci flat metric. The solution \eq{MNsol} gives the solution of \cite{Chamseddine:1997nm,Maldacena:2000yy} and has a finite size $S^3$.  One should define the parameter which interpolates between these two solutions as the size of the $S^3$ in string units
\be
U\sim \frac{R(S^3)}{g_s},
\ee
$U=0$ is the CV/MN solution and $U\ra \infty$ is the solution with the Ricci flat metric on the deformed conifold. This the bottom line in figure 1.

One thing to note from \eq{BBsol} is that the holomorphic three form (suitably rescaled by the warp factor) is invariant along this family \cite{Maldacena:2009mw}. This is consistent with the fact that this family is a realization in IIB string theory of Vafa's geometric transition \cite{Vafa:2000wi} and only depends on the ``Kahler" moduli, not the complex structure moduli. The quotation marks are necessary here because the IIB background is not Kahler away from $U=\infty$. In string frame, the equations for a system of just $(g,\vphi,F_3)$ is given by \cite{Strominger:1986uh, Hull:1986kz,Gauntlett:2003cy}
\bea
d(e^{A}\Om)&=&=0 \label{F31} \\
d(e^{2A}J)&=&*e^{4A} F_3,  \label{F32}\\
d(J\w J)&=&0 \label{F33}, \\
2A&=&\vphi . \label{F34}
\eea
We see that \eq{F31} is solved somewhat trivially along the whole family since $e^{A}\Om$ is invariant. It is also interesting that the $F_3$ flux is invariant along this family but the deeper reason for this is not clear.
\begin{figure}[bth!]
\begin{center}
\includegraphics[width=12cm,height=9cm]{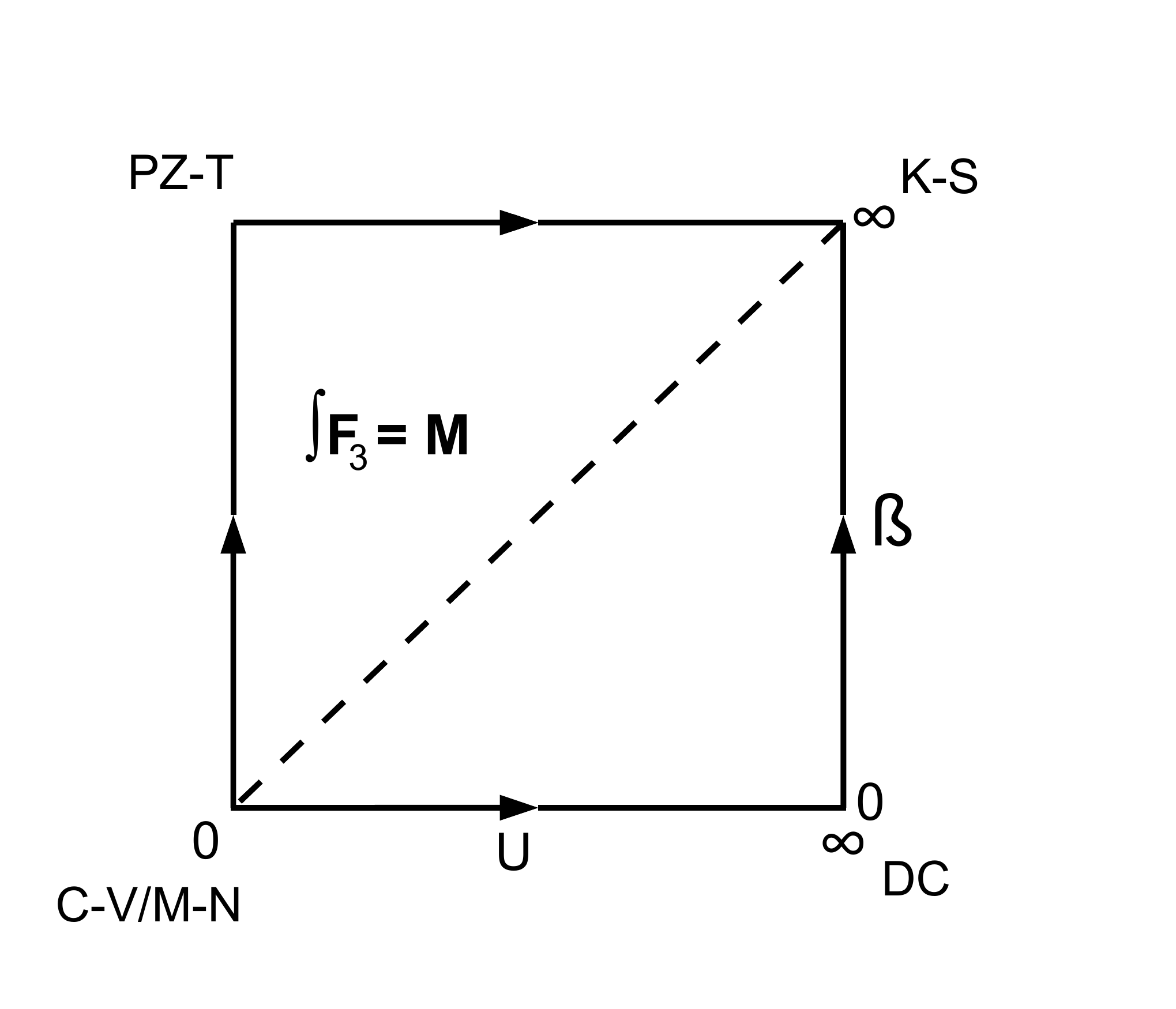}
\caption{The parameter space of the deformed conifold in IIB.}
\label{DefCon}
\end{center}
\end{figure}

As emphasized in \cite{Maldacena:2009mw} there is another parameter in this moduli space of solutions. The full solution space is given by
\bea
ds_{str}^2&=& h^{-1/2} ds_4^2 + \frac{M h^{1/2}}{\cosh\, \beta} ds_{M_6}^2 \\
h&=& 1+ \cosh \, \beta (e^{2(\vphi-\vphi_\infty)} -1) \\
F_3&=& \frac{\alpha' M}{4} \om_3 \\
H_3&=& -\tanh\, \beta \frac{e^{\vphi_\infty} M \alpha'}{4} e^{-2(\vphi-\vphi_\infty)} *_6 \om_3 \\
F_5&=& -\tanh \, \beta e^{-\vphi_\infty}(1+*_{10}) \vol_4 \w d(h^{-1}).
\eea
Figure 1. represents the full moduli space of $(U,\beta)$. 

The $U\ra \infty, \beta \ra \infty$ limit is the solution of Klebanov and Strassler (KS) and the dashed line in figure 1. represents a one-parameter interpolation between the KS background and the CV/MN background which In \cite{Butti:2004pk}  was proposed to be the supergravity dual to the baryonic branch of the Klebanov-Strassler background \cite{Gubser:2004qj}. However as was empahasized in  \cite{Dymarsky:2005xt, Maldacena:2009mw} the top line of this diagram, namely with $\beta\ra \infty$ and $U=[0,\infty]$ has the asymptotic behavior consistent with  the baryonic branch. As was described in some detail in \cite{Maldacena:2009mw}, the solution at the top left corner in figure 1, with $U=0,\beta=\infty$ is a little hard to describe in closed form but it has a certain region in which it approximates the solution of Pando-Zayas and Tseytlin \cite{PandoZayas:2000sq}. 

It is interesting to note that the operator which deforms the KS solution along the vertical $\beta$ direction is a dimension eight operator which is $SU(2)^2\times U(1)\times \ZZ_2$ invariant. From the table of modes collated in \cite{Bena:2009xk} we see that there is only one such operator. Indeed, using the dimensions of operators from the conformal point \cite{Klebanov:1998hh} we see that this operator must be $\Tr F^4$. This operator is clearly irrelevant and changes the UV definition of the gauge theory, thus changes the asymptotic behaviour of the warp factor.  Interestingly, the vev for this operator is related to the addition of anti-D3 branes into the KS background.

\section{The Resolved Conifold in Type IIA Supergravity} \label{IIAres}
We now turn to the mirror picture of the previous section, namely the resolved conifold in IIA.
There exists a family of solutions of IIA supergravity on the resolved conifold found by Brandhuber \cite{Brandhuber:2001kq} which we propose is the mirror to the $\beta=0$ limit of the solutions in the previous section. In fact these solutions were found in M-theory, where the computations are simplified significantly as we now review.

In M-theory on a seven-manifold with $G^{(4)}=0$, there are just two nontrivial spinor bilinears one can construct,
a three form and a four form  \cite{Candelas:1984yd}
\bea
\Phi_{abc}&=&i\tha^{\dagger}\gam_{abc} \tha, \non \\
(*\Phi)_{abcd} &=&\tha^{\dagger}\gam_{abcd} \tha \non
\eea
and the conditions for $\N=1$ supersymmetry in four dimensions are simply
\be
d \Phi=0,\ \ \ d*\Phi=0. \label{G2conditions}
\ee
These conditions imply that there exists a connection of $G_2$ holonomy. The simplicity  of working in M-theory with $G^{(4)}=0$ is that one need just make an ansatz for the three form $\Phi$,
since the metric and thus the four form $*\Phi$ can be constructed from $\Phi$ using 
\bea
g_{ij}&=& (\det s_{ij})^{-1/9}s_{ij}, \non \\
s_{ij}&=&-\frac{1}{144} \Phi_{im_1 m_2 }\Phi_{jm_3 m_4 }\Phi_{m_5 m_6 m_7} \eps^{m_1\ldots m_7},\ \ \eps^{1234567}=1. \non
\eea

When searching for supergravity solutions, common sense tends to indicate that 
one should (at least initially) restrict attention to cohomogeneity one solutions since
this will result in a system of O.D.E.'s not P.D.E.'s. In the case of a seven manifold,
we will impose the continuous symmetry group
\be
SU(2)_{L,1}\times SU(2)_{L,2}\times U(1)_{R,D} \label{sym1}
\ee
where $SU(2)_{L,1}$ and $SU(2)_{L,2}$ are parameterized by left invariant one forms $\sig_i$ and $\Sig_j$ 
and $U(1)_{R,D}\subset SU(2)_{R,1}\times SU(2)_{R,2}$ is diagonally embedded.
The most general invariant three form is then
\bea
\Phi&=& p\, \sig_1\w \sig_2\w \sig_3 + q\, \Sig_1 \w \Sig_2\w \Sig_3 \non \\
&&+ a(r)\blp \sig_1\w \Sig_1 + \sig_2 \w \Sig_2\brp + b(r) \sig_3 \w \Sig_3 \label{andyphi}
\eea
which depends on merely 2 functions, one of which can be fixed by redefining the radial co-ordinate! The solution we will focus on breaks the $\ZZ_2$ symmetry which exchanges $S^3_1\ra S^3_2$, it has $(p,q)=r_0^2(1,0)$.
One then finds that $d*\Phi=0$ implies a single second order O.D.E.
\bea
0&=&4 a' b'\Blp ab(b+r_0^2)a'+(b^3-a^2(r_0^2+2b))b' \Brp+b(b^3-4a^2 (r_0^2+b))(a'b''-a''b').
 \label{andyeq}
\eea
The whole family of solutions of \cite{Brandhuber:2001kq}, when reduced to type IIA 
using the standard formula
\be
ds_{11}^2=e^{-2\vphi/3} ds_{10}^2 + e^{4\vphi/3} (d\psi+C_{1}),
\ee
is given by\footnote{we have chosen the radial co-ordinate $b(r)=r^2/6$ to agree with the conventional radial co-ordinate on the resolved conifold.}
\bea
ds_{10}^2&=& e^{-2\vphi/3} ds_4^2 + \half ds_{M_6}^2, \non \\
F_2 &=& N(\sig_1\w \sig_2 + \Sig_1\w\Sig_2)+ d\Blp  \frac{6r_0^2}{6r_0^2 +r^2} (\sig_3 + \Sig_3) \Brp, \\
e^{2\vphi}&=& \frac{r^3(r^2+6r_0^2)}{72a'}\sqrt{\frac{(r^2+6r_0^2)}{ 144 (r^2+6r_0^2)a^2-r^6}} \non
\eea 
where the frames on $M_6$ are given by
\bea
F_1&=& \cosh(B_1)  E_1 + \sinh(B_1) \Ebar_2, \non \\
F_2&=&  \cosh(B_1) E_2 + \sinh(B_1)\Ebar_1, \\
F_3&=&  \cosh(B_2) E_3 + \sinh(B_2) \Ebar_3 \non 
\eea 
and the complex frames $E_i$ are those of the Ricci flat metric on the resolved conifold
\bea
E_1&=&\frac{r}{\sqrt{6}} (\sig_1+i\sig_2),  \non \\
E_2&=& (r_0^2+\frac{r^2}{6})^{1/2}(\Sig_1+i\Sig_2), \\
E_3&=& \kappa^{-1/2} dr+i \frac{r\kappa^{1/2}}{3} (\sig_3 + \Sig_3).\non
\eea
The functions are given by
\bea
\kappa&=& \frac{r^2 +9 r_0^2}{r^2 + 6 r_0^2 } \non \\
\sinh\, 2 B_1&=& -\frac{r^3}{\sqrt{144a^2 (r^2+6 r_0^2)-r^6}}  \\
e^{ 2 B_2}&=& -\frac{\sqrt{144 a^2 (r^2+6r_0^2)- r^6 }}{8ra' \kappa \sqrt{r^2+6r_0^2}}\non 
\eea

An obvious solution to \eq{andyeq} is given by 
\be
BS/GPP:\ \ a=b\, ,
\ee 
where in fact the symmetry group \eq{sym1} is enlarged to 
\be
SU(2)_1\times SU(2)_2\times SU(2)_D
\ee
and this is the solution originally found in \cite{bryant, Gibbons:1989er}. There is then another explicit solution to \eq{andyeq} which is 
\be
RC:\ \ \lim_{U\ra \infty}\ a=\frac{1}{U}\frac{r^2}{6},\ \ b=\frac{r^2}{6}.
\ee 
This limit must be taken carefully, the asymptotic value of the dilaton is constant and the radius of the $S^2$ is large
\be
U= \frac{r_0}{e^{\phi_\infty}} >>0,\ \ \ r_0>>0,\ \ \ e^{\phi_\infty}={\rm constant}.
\ee
 
\begin{figure}[bth!]
\begin{center}
\includegraphics[width=12cm,height=5cm]{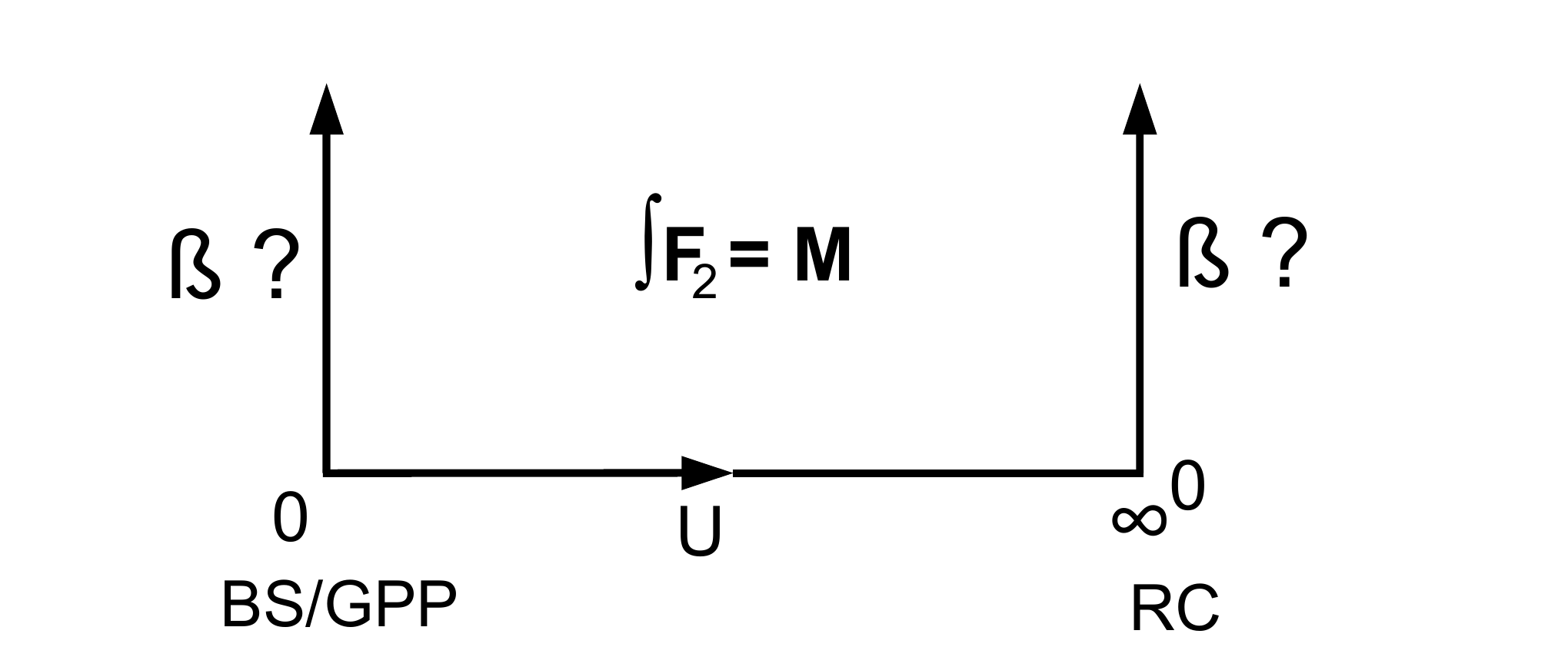}
\caption{The proposed parameter space of the resolved conifold in IIA. The bottom line ($\beta=0$) corresponds to the solutions found in \protect{\cite{Brandhuber:2001kq}} and $\beta\neq 0$ is the subject of the current work.}
\label{resConfig}
\end{center}
\end{figure}

We notice a striking analogy between these solutions and the IIB solutions presented in the previous section. Here the symplectic form  and flux  $(J,F_2)$ which are invariant along the family while $(\Om,\vphi)$ vary, where previous it was $(\Om,F_3)$ which were invariant while $(J,\vphi)$ varied. One can also see the invariance of the symplectic form $J$ from the fact that it is the dimensional reduction of \eq{andyphi} along the vector field dual generated by $\del_\psi-\del_{\tpsi}$ gives
\be
J=\half (p + b ) \sig_1\w \sig_2  +\half (q - b ) \Sig_1\w \Sig_2  +\half b'dr\w  (\sig_3+\Sig_3)\non 
\ee
and this does not depend on $a(r)$ only on $b(r)$. Choosing $b(r)=r^2/6$ and $(p,q)=r_0^2(1,0)$  gives $J=J_{RC}$. This family of supergravity solutions in IIA displays all the features of the geometric transition as a smooth process in flux backgrounds which were emphasized in the IIB case in \cite{Maldacena:2009mw}. This is somewhat at odds with the view presented in \cite{Atiyah:2000zz} where it was proposed that the transition is a discrete process obtained as the reduction of the BS/GPP solution on two different $U(1)$ fibers. 

It is natural to wonder if there exists another parameter in these IIA solutions which would be mirror to $\beta$ in the IIB solutions and the remainder of this paper is devoted to this question. Since the IIB $\beta$ parameter in Table \ref{DefCon} was shown to be generated by a certain duality transformation in \cite{Maldacena:2009mw} it would be nice if one could use a different duality transformation in this IIA setting. One way to attempt this is to T-dualize four times in space-time to turn the D6 branes into Euclidean D2 branes then lift to M-theory and use a diffeomorphism on a $T^2$. The essential problem with this is that any rotation of a $T^2$ requires both circles to be finite in the UV, which is not the case. As such, probably the correct duality transformation requires mirror symmetry, which as discussed above is hard to perform explicitly, so here we will just use the supergravity ansatz method.

\section{IIA on the Resolved Conifold with General Fluxes}
Motivated by the previous discussion we now search for solutions of massive IIA supergravity with a four dimensional Minkowski component and $SU(2)_{L,1}\times SU(2)_{L,2}\times U(1)_{R,D}$ symmetry on the internal space. All the main calculations have been relegated to the appendix, we summarize our findings in this section.

When solutions of IIA with flux are lifted to $d=11$ supergravity, many aspects of the backgrounds simplify considerably. For instance, the solutions found by Brandhuber which we reviewed in section \ref{IIAres}. involve a single nontrivial function whereas the ansatz with the same symmetries in IIA involves seven functions  and are somewhat more complicated to obtain. Given this our strategy will be the following: we will search for the most general solution of M-theory with our symmetries \eq{sym1} and allow for nontrivial $G^{(4)}$ flux. Then we will look for solutions of massive IIA supergravity with the same symmetries however we will find fairly quickly that the mass parameter is forced to vanish. As such we conclude that the solutions we found in M-theory are the most general possible solutions.

\subsection{Supersymmetry in M-theory and IIA}

In type II supergravity reduced to four dimensions preserving $\N=1$ supersymmetry, the most general possibility for the supersymmetry parameters is an $SU(2)$ structure. It is instructive to think of the $SU(2)$ as being the intersection of two $SU(3)$'s, one for the left movers (on the string worldsheet) and one for the right movers and these are often referred to as $SU(3)\times SU(3)$ structures. If these two $SU(3)$ structures are aligned at some subvariety of the internal space, then the $SU(2)$ structure becomes an $SU(3)$ structure at there. Such cases are called {\it dynamic} $SU(2)$ structures and in terms of generalized complex geometry \cite{Hitchin:2004ut,Gualtieri:2003dx} there is {\it type change}. If the two $SU(3)$ structure are never parallel, then the $SU(2)$ structure is called {\it static}. If the two $SU(3)$ structures are parallel everywhere, then such backgrounds are simply called $SU(3)$ structures. 

There is a distinct asymmetry between $SU(3)$ structures in IIA and IIB. Whereas in IIB, $SU(3)$ structures allow for non-trivial Ramond fluxes of all degrees $(F_1,F_3,F_5)$ as well as $(\vphi,H_3)$, the allowed fluxes for $SU(3)$ structure solutions of IIA with $\RR_{1,3}$ component are very limited. One possibility is for an NS solution of the form \cite{Strominger:1986uh, Hull:1986kz} and the other possibility is to have non-trivial profiles for only $(g,\vphi,F_2)$.  The former case lifts to M-theory as a solution with $G^{(4)}$ while the second case lifts to a solution of M-theory on a seven manifold with $G_2$ holonomy and $G^{(4)}=0$. Allowing for $F_0\neq 0$ in IIA $SU(3)$ structure solutions generates a four dimensional cosmological constant and it seems unlikely that mirror symmetry exchanges Minkowski vacua with AdS vacua.

However, it is not entirely clear that mirror symmetry with flux (if it exists) maps one $SU(3)$ structure solution to another $SU(3)$ structure solution. For instance it would be quite reasonable to expect an $SU(3)$ structure solution to be mapped to a static $SU(2)$ structure solution although it would be quite surprising if a solution with type change was mapped to a solution without type change. It is likely that restoration of a symmetry between solutions of string theory in IIA and IIB requires an understanding of non-geometric backgrounds, while
these are interesting issues we cannot resolve them here. We will merely examine the example of supergravity solutions with co-homogeneity one symmetry groups.

\subsection{The Ansatz and Solution}

Our  metric ansatz is in $d=11$ supergravity is
\be
ds_{11}^2=e^{2\Delta} ds_{1,3}^2 + ds_7^2
\ee
with the frames on the internal seven manifold given by
\bea
e_1= A_2(\Sig_1-A_4\sig_1),&& e_2= A_1\sig_1,  \non \\
e_3= A_2(\Sig_2-A_4\sig_2),&& e_4= A_1\sig_2,  \non \\
e_5= A_5(\Sig_3-A_7\sig_3),&& e_6= A_6\sig_3,  \non \\
e_7=A_3 dr. \non
\eea
We will find the four form flux $G^{(4)}$ algebraically in terms of metric data so there is no need to make an ansatz for it.
We find that solving the supersymmetry equations gives all the metric functions to be
\bea
e^{6\Delta}&=& \frac{4r^6(r^2(1-g_3^2/4)+6 r_0^2)}{\blp 4r^6(r^2(1-g_3^2/4)+6 r_0^2)-144 c^2\brp}  \non \\
A_1^6&=& 4 (r^2(1-g_3^2/4)+6 r_0^2)) \frac{\blp 4r^6(r^2(1-g_3^2/4)+6 r_0^2)-144 c^2\brp (-g_3+r g'_3)}{2^7 3^2\, r^6 g_3^3 } \non \\
A_2^6&=&  \frac{r^6}{\blp r^2(1-g_3^2/4) +6r_0^2\brp^3}  A_1^6  \non \\
A_3&=& -\frac{2 e^{3\Delta}}{r^2} \frac{A_2^2}{A_5} (r^2+8 r_0^2)^{2/3} \non \\
A_4&=&\frac{g_3}{2} \non \\
A_5&=& \frac{A_2^2}{1-r (\log\,g_3)'/2}\non \\
 A_6 &=&\frac{r^2}{6} \frac{1}{e^{3\Delta} A_2^2} \non \\
 A_7&=& 1-\frac{g_3^2}{2}\non
\eea
where the function $g_3$ is defined to be
\be
g_3=\frac{A_2 A_6}{A_1 A_5} \non
\ee
and $(c,r_0)$ are integration constants. We redefine $g_3$ to be 
\be
g_3=\frac{r^2}{6a} \non 
\ee
then $a(r)$ satisfies the non-linear equation
\bea
0&=&9 a' \blp-16 c^2-r^8+16 r^2 a ((18 r_0^2+5 r^2) a-r (6 r_0^2+r^2) a')\brp\non \\
&&+r \blp 144 c^2+r^8-144 r^2 (6 r_0^2 +r^2) a^2\brp a''. \label{myeq}
\eea
The function $a(r)$ agrees with the function $a(r)$ in \eq{andyeq} if ones defines the function $b(r)$  to be $b(r)=r^2/6$.

These solutions are in general singular when 
\be
0=r^6(r^2(1-g_3^2/4)+6 r_0^2)-36 c^2
\ee
and this appears to be a particularly bad type of singularity where the warp factors diverges and the two internal $S^3$'s shrink to zero size. When $c=0$, the whole family of solutions reduce to the regular solutions discussed in section \ref{IIAres} and $r_0$ is the radius of the finite $S^2$ (in IIA) at $r=0$.

\subsection{The $SU(2)^3$ Invariant Solution}
One solution to \eq{myeq} is
\be
a(r)=r^2/6.
\ee
In this case we have
\bea
A_6=A_1,\ \ && A_5=A_2 \\
A_7=A_4,\ \ &&  g_3=1. \non
\eea
and the $U(1)_{R,D}$ is enhanced to $SU(2)_{R,D}$. This solution reduces to the BS/GP solution when $c=0$.

We find that
\bea
A_1^6 &=&  \frac{9(r^2+8r_0^2)^2}{16 r^4 } \Blp \frac{r^4}{36} - \frac{4c^2}{ 3 r^2(r^2+ 8r_0^2)}  \Brp \\
A_2^6 &=& \frac{4r^2}{3(r^2+8r_0^2)} \Blp \frac{r^4}{36} - \frac{4c^2}{ 3 r^2(r^2+ 8r_0^2)}  \Brp \\
e^{6\Delta}&=& \frac{r^4}{36}  \Blp \frac{r^4}{36} - \frac{4c^2}{ 3 r^2(r^2+ 8r_0^2)}  \Brp^{-1}
\eea
so this solution is singular when
\be
c^2=\frac{r^6}{48}(r^2+8 r_0^2).
\ee
The corrections when $c\neq0$ are very subleading in the UV $r\ra \infty$ limit. More generally the UV behavior is more complicated since then $g_3$ is a function of $c$.

\section{Conclusions: The Missing Mirrors}

So while we have found a one dimensional family of solutions which extends those found in \cite{Brandhuber:2001kq}, they appear to be badly singular. It is not inconceivable that these singularities are somehow repaired in string theory however since we have considered the most general supergravity ansatz, it must be some stringy (as opposed to supergravity) effects which resolve them. This is unlike the singularity of \cite{Klebanov:2000nc} which is repaired within supergravity \cite{Klebanov:2000hb}.

Having explored the most general supergravity ansatz for a putative mirror to the IIB solutions of \cite{Butti:2004pk} we must confront the fact that the mirror may be missing, much like the third T-dual of $T^3$ with worldvolume $H_3$ flux is missing geometrically \cite{Bouwknegt:2004ap, Hull:2004in, Shelton:2005cf}. In a sense this may not be such a surprise since as discussed at the above, there is a distinct asymmetry between $SU(3)$ structures in IIA and IIB supergravity. It was realized in \cite{Shelton:2005cf} that to formulate a low energy description of string compactifications which is duality symmetric appears to require additional structures which do not admit a geometric interpretation. A more sophisticated version of the same ideas appeared in \cite{Grana:2006hr}.  However what we have found here is somewhat more bizarre since it is not magnetic flux which is proving to be problematic. In section \ref{IIAres} we proposed a mirror to the magnetic $F_{3}$ flux, namely the magnetic $F_2$ flux, it is the additional non-topological $H_3$ and $F_5$ flux which appears to be problematic in terms of finding the mirror dual.

We consider this calculation as shining a light on a more general sort of non-geometricity than that seen in earlier works which largely deal with tori and monodromy around non-contractible 1-cycles.  
It would be extremely interesting if some sort of explicit solution for these mirrors could be found which involves some of the ideas of non-geometric string backgrounds advocated in \cite{Halmagyi:2008dr, Halmagyi:2009te} where a certain bi-vector coupling on the string worldsheet was used.  One strategy might be to extend the nice four dimensional theory of \cite{Koerber:2007xk} to include non-geometric fluxes and then attempt to solve explicitly for some sort of ansatz which employs the same symmetries we have applied in this work. 

\vskip 1cm
\noindent {\bf Acknowledgements:}
I would like to thank Iosif Bena, Nikolay Bobev, Jerome Gauntlett, Mariana Gra\~na, Igor Klebanov, Juan Maldacena, Michela Petrini and Alessandro Tomasiello for discussions. This work was supported in part by the DSM CEA-Saclay, by the ANR grants BLAN 06-3-137168 and JCJC ERCS07-12 and by the Marie Curie IRG 046430.

\begin{appendix}

\section{Conventions}

We will use the following conventions for the $SU(2)_L$-invariant one forms 
\bea
d\sig_i&=& \half \eps_{ijk} \sig_j\w \sig_k \non \\
d\Sig_i&=& \half \eps_{ijk} \Sig_j\w \Sig_k. \non \\
\eea
It is very common to use $e_i$ and $\eps_i$ however since we use $e_i$ for frames and $\eps_i$ for spinors, we have chosen to use these conventions for the invariant one forms. We will denote the two sets of Euler angles as $(\tha,\phi,\psi)$ for the $\sig_i$ and $(\ttha,\tphi,\tpsi)$ for the $\Sig_i.$

\section{$G_2\times G_2$ Structure solutions in $d=11$ Supergravity}

\subsection{The Gravitino Variation}

The gravitino variation of eleven dimensional supergravity is 
\be
\delta \Psi_A = \hat{\nabla}_A \hat{\eps} + \frac{1}{288} \Blp G_{BCDE}\hgam^{BCDE}_{\ \ \ \ \ \ A} -8 \delta^{B}_{A} \hgam^{CDE} \Brp \hat{\eps}. \label{11grav}
\ee
where the hatted objects here denoted eleven dimensional objects. The most general spinor ansatz which preserves $\N=1$ supersymmetry in four dimensions is
\be
\hat{\eps} = \zeta_+ \otimes \tha_+ + \zeta_- \otimes \tha_-.
\ee
Here we have that $\zeta_+ =\zeta_-^*$ is a four dimensional Weyl spinor of positive chirality and  
\be
\tha_+=\eps_1 + i \eps_2 \label{}
\ee
is a complex seven dimensional spinor. In general this spinor ansatz appears to have too many degrees of freedom however $\eps_1$ and $\eps_2$ will not be independent. 

The metric and four form ansatz is
\bea
ds_{11}^2&=&e^{2\Delta} ds_{1,3}^2 + ds_7^2 \non \\
G^{(4)}&=& \frac{1}{4!}G^{(4)}_{ijkl} dx^i\w dx^j \w dx^k \w dx^l \non
\eea
where $(i,j,k,l)$ are co-ordinate indices on the internal space. Allowing for flux proportional to the volume form on the four dimensional component of space allows for a cosmological constant, we will not consider that possibility here.

With this ansatz, the eleven dimensional gravitino variation  \eq{11grav} reduces to an external (algebraic) and internal (differential) component:
\bea
0&=&\Blp\pm \half (\del_c \Delta) \gam^c + \frac{1}{288} G^{(4)}_{bcde} \gam^{bcde} \Brp \tha_\pm,  \label{susyext} \\
\nabla_a \tha_{\pm} &=& \mp \frac{1}{288} \Blp G^{(4)}_{bcde} \gam^{bcde}_{\ \ \ \ a} - 8 G^{(4)}_{abcd} \gam^{bcd}\Brp \tha_\pm \label{susyint}
\eea
where now the unhatted gamma matrices are seven dimensional ones and the indices $(a,b,c,d)$ are internal frame indices.

\subsubsection{The Differential Forms}

The by now standard procedure \cite{Gauntlett:2002fz}, is to use the spinor equations \eq{susyext} and \eq{susyint} to pass to a set of equations for the differential forms constructed as spinor bi-linears. The alternative approach to finding supersymmetric backgrounds is to make an ansatz for the spinor in addition to the bosonic fields and solve \eq{11grav} directly \!\!\!\!
\footnote{Such an approach is most definitely equivalent in content but perhaps more computationally intensive, a small representative set of examples with non-trivial projectors on the spinors might be \cite{Pilch:2003jg, Bena:2004jw,Halmagyi:2005pn}.}.

The advantage of constructing the differential forms and computing the differential equations they satisfy is that with this finite amount of work (which can then be universally applied to all ansatze) one alleviates the need to solve for the spinor fields, thus reducing the overall number of functions in the ansatz. In addition, it seems promising that formal properties of supergravity backgrounds can be better understood by studying the differential forms \cite{Strominger:1986uh,Fu:2006vj, Gauntlett:2003cy, Grana:2005sn, Tomasiello:2007zq}.
The equations we adopt here could be lifted from the nice paper \cite{Lukas:2004ip} where in fact the more general situation with a four dimensional cosmological constant was considered\!\!
\footnote{The supersymmetry conditions for the reduction of M-theory on a seven manifold with background flux was also considered in \cite{Kaste:2003zd,Dall'Agata:2003ir,Gauntlett:2004zh}. These papers consider a spinor ansatz which is not sufficiently general for our purposes since there $|\eps_1|=|\eps_2|$ while here we allow for the limit $\eps_2\ra 0$ since this corresponds to $G^{(4)}\ra 0$ and the restoration of $G_2$ holonomy.}. 

So we consider the following collection of differential forms
\bea
\Xi_{a_1\ldots a_m}&=&\tha_+^{\dagger} \gam_{a_1\ldots a_m}\tha_+, \label{Xi}\\
\Xit_{a_1\ldots a_n}&=&\tha_-^{\dagger} \gam_{a_1\ldots a_n}\tha_+ \label{Xit}
\eea
and relate the two spinors $\eps_i$ by
\be
\eps_2=\lam \gam_{12}\eps_1
\ee
with $\lambda$ an arbitrary function. The limit $\lambda\ra 0$ will coincide with $\{ G^{(4)}\ra 0,\Delta\ra 0\}$ and thus such a seven dimensional metric admits a connection with $G_2$ holonomy.

In terms of the frames of the seven dimensional metric $\{ e_i\},\ i=1,\ldots ,7$ we define certain fundamental forms
\bea
V&=&e_7 \non \\
J&=&e_1\w e_2 + e_3\w e_4 + e_5\w e_6 \non \\
\Psi&=& \Psi_+ + i \Psi_- \non \\
&=& (e_1+ ie_2)\w (e_3+ie_4)\w(e_5+ie_6) \non
\eea
and express \eq{Xi} and \eq{Xit} in terms of them
\bea
\Xi&=& (1+\lam^2) |\eps_1|^2, \non \\
\Xit&=&  (1-\lam^2) |\eps_1|^2,\non \\
\Xi^{(1)}&=&  \sqrt{\Xi^2-\Xit^2} V,\non \\
i\Xi^{(2)}&=& \sqrt{\Xi^2-\Xit^2} J, \non \\
i\Xi^{(3)}&=&\Xit \Psi_- -\Xi J\w V, \non \\
\Xit^{(3)}&=&\sqrt{\Xi^2-\Xit^2}  \Psi_+ + i\Blp \Xi\Psi_-  - \Xit J\w V \Brp  \non. 
\eea
The following differential equations are computed by first evaluating the exterior derivative of a given $\Xi^{(i)}$
or $\Xit^{(j)}$ using the differential part of the spinor variation \eq{susyint} and then using the algebraic part \eq{susyext} to replace certain flux terms. For this laborious calculation this author recommends consulting the nice summary of formulas in \cite{Candelas:1984yd} and the computer package \cite{Gran:2001yh}. The result is
\bea
 d\Blp e^{-\Delta} \Xi\Brp  = 0,&&
 d\Blp e^{2\Delta} \Xit\Brp  = 0 \\
 d\Blp e^{\Delta} \Xi^{(1)}\Brp  &=& 0   \\
 e^{-3\Delta}d\Blp e^{3\Delta }\Xi^{(2)}\Brp &=& i \Xi (*G) \\
 e^{-5\Delta}d\Blp e^{5\Delta }\Xi^{(3)}\Brp &=& -i \Xi^{(1)} \w (*G) \\
e^{-2\Delta}d\Blp e^{2\Delta }\Xit^{(3)}\Brp &=& -\Xit  G \\
 e^{-\Delta}d\Blp e^{\Delta }\Xit^{(4)}\Brp &=& G\w \Xi^{(1)}
\eea
In terms of the fundamental forms this gives
\bea
 d\Blp e^{-\Delta} \Xi\Brp  = 0, \label{form0}&&
 d\Blp e^{2\Delta} \Xit\Brp  = 0, \label{form1}  \\
 d\Blp e^{\Delta} \sqrt{\Xi^2-\Xit^2} V \Brp  &=& 0,\label{form2}    \\
 e^{-3\Delta}d\Blp e^{3\Delta }\sqrt{\Xi^2-\Xit^2} J\Brp &=& - \Xi (*G), \label{form3}  \\
 e^{-5\Delta}d\Blp e^{5\Delta }( \Xit \Psi_- -\Xi J\w V) \Brp &=&\sqrt{\Xi^2-\Xit^2}\, V \w (*G), \label{form4}  \\
e^{-2\Delta}d\Blp e^{2\Delta }\sqrt{\Xi^2-\Xit^2}  \Psi_+\Brp &=& -\Xit  G,  \label{form5}  \\
e^{-2\Delta}d\Blp e^{2\Delta }\Blp \Xi\Psi_-  - \Xit J\w V \Brp \Brp&=& 0, \label{form6}  \\
 e^{-\Delta}d\Blp e^{\Delta }\Xit^{(4)}\Brp &=& \sqrt{\Xi^2-\Xit^2}\, G\w V \label{form7} .
\eea
This system is most certainly overcomplete.

It is important to obseve that the limit $\lam\ra 0$ reduces to the more familiar case \eq{G2conditions}. In this limit $\Xi,\Xit \ra 1$ which implies that $\Delta$ is constant. In addition $\Xi^{(1)},\Xi^{(2)}\ra 0$ and $\Xi^{(3)}\ra -\Xit^{(3)}\ra \Phi$.

\subsection{Ansatz For the Metric}
Since the four form flux $G^{(4)}$ is given algebraically by \eq{form5}, we only need to make an ansatz for the metric. We will take:
\bea
e_1= A_2(\Sig_1-A_4\sig_1),&& e_2= A_1\sig_1,  \non \\
e_3= A_2(\Sig_2-A_4\sig_2),&& e_4= A_1\sig_2,  \non \\
e_5= A_5(\Sig_3-A_7\sig_3),&& e_6= A_6\sig_3,  \non \\
e_7=A_3 dr \non.
\eea
This in general breaks the $\ZZ_2$ which exchanges $\Sig_i\lra \sig_i$ and preserves $U(1)_{R,D} \subset SU(2)_{R,D}$.
So now with these frames the spinor bilinears are
\bea
J&=&e_1\w e_2 + e_3 \w e_4 + e_4 \w e_6 \non \\
&=&-A_1A_2( \sig_1\w  \Sig_1+ \sig_2\w  \Sig_2 ) -A_5 A_6 \sig_3\w \Sig_3,\non \\
\Psi_+&=& \blp 2A_1 A_2 A_4 A_6+A_1^2 A_5 A_7 - A_2^2 A_4^2 A_5 A_7 \brp \sig_{123} \non \\
&&  +A_2^2A_5 \Sig_{123} \non \\
&&+\Blp- A_1 A_2 A_6 + A_2 ^2 A_4 A_5 A_7  \Brp (\sig_{23}\Sig_1 + \sig_{31}\Sig_2)  \non \\
&& +(-A_1^2 A_5 + A_2^2 A_4^2 A_5)\sig_{12}\Sig_3  \non \\
&&- A_2^2 A_4 A_5 \blp \sig_1\Sig_{23} + \sig_2 \Sig_{31}\brp \non\\
&&- A_2^2 A_5 A_7 \sig_3 \Sig_{12} ,\non \\ 
\Psi_-&=&\blp -A_1^2A_6+A_2^2A_4^2A_6+2A_1A_2 A_4 A_5 A_7 \brp \sig_{123} \non \\
&& -(2A_1A_2 A_4 A_5)\sig_{12}\Sig_3  \non \\
&&- (A_2^2 A_4 A_6 + A_1 A_2 A_5 A_7)(\sig_{23}\Sig_1 + \sig_{31}\Sig_2)  \non   \\
&&+ A_2^2 A_6 \sig_3 \Sig_{12} \non \\
&&+ A_1 A_2 A_5(\sig_1\Sig_{23} + \sig_2 \Sig_{31}). \non
\eea

\subsection{Solving}
Without loss of generality we can take 
\bea
\Xi&=&e^{\Delta} \non \\
\Xit&=&e^{-2\Delta} \non \\
\Rightarrow \sqrt{\Xi^2 - \Xit^2}&=& e^{\Delta}\sqrt{1-e^{-6\Delta}} \non
\eea
\subsubsection{$d\Blp \Xi^{(3)} \Brp$}
So we start with
\bea
\eq{form6}:\ \ 0&=&d\Blp  e^{3\Delta} \Psi_- - J\w V\Brp \non \\
&=& \blp e^{3\Delta}  \blp -A_1^2A_6+A_2^2A_4^2A_6+2A_1A_2 A_4 A_5 A_7  \brp \brp 'dr\w \sig_{123} \non \\
&&+\Blp  e^{3\Delta} (-2A_1A_2 A_4 A_5 + A_2^2 A_6)  \Brp \sig_{12}\Sig_{12} \non \\
&&+ \Blp  e^{3\Delta} \blp -A_2^2 A_4 A_6 -A_1 A_2 A_5 A_7 +A_1 A_2 A_5\brp\Brp (\sig_{23}\Sig_{23} + \sig_{13}\Sig_{13}) \non \\
&&+\Blp -\blp e^{3\Delta}(A_2^2 A_4 A_6 + A_1 A_2 A_5 A_7) \brp' -A_1A_2A_3 \Brp dr\w(\sig_{23}\Sig_1 + \sig_{31}\Sig_2)   \non \\
&&+ \Blp -\blp e^{3\Delta}2A_1 A_2 A_4 A_5\brp'  - A_3 A_5 A_6 \Brp dr\w\sig_{12}\Sig_3  \non \\
&&+ \Blp  \blp e^{3\Delta}A_1 A_2 A_5 \brp' + A_1A_2A_3  \Brp dr\w(\sig_{1}\Sig_{23} + \sig_{2}\Sig_{13})   \non \\
&&+ \Blp\blp e^{3\Delta}A_2^2 A_6 \brp' + A_3 A_5 A_6\Brp dr\w\sig_{3}\Sig_{12}  \label{dXi3} 
\eea
First, we get the algebraic constraints
\bea
A_2 A_6 &=& 2 A_1 A_4 A_5 \non \\
A_1 A_5 (1-A_7) &=& A_2 A_4 A_6 \non \\ 
\Rightarrow A_7 &= 1- 2 A_4^2 \non \\
{\rm and}\ \ \ \ A_4&=& \frac{A_2 A_6}{2 A_1 A_5} \non
\eea
This means we have solved for $\{A_4,A_7\}$ in terms of the other functions. When we use these algebraic relations above, we get just two equations
\bea
\blp e^{3\Delta} A_2^2 A_6\brp' &=& - A_3 A_5 A_6 \non \\
\Rightarrow \Blp \log \blp e^{3\Delta} A_2^2 A_6\brp\Brp' &=& - e^{-3\Delta}\frac{A_3 A_5}{A_2^2} \label{Xi3eq1} \\
\blp e^{3\Delta} A_1 A_2 A_5\brp' &=& - A_1 A_2 A_3 \non \\
\Rightarrow \Blp \log \blp e^{3\Delta} A_1 A_2 A_5\brp\Brp' &=& - e^{-3\Delta}\frac{A_3}{A_5} 
\eea
Looking ahead to the sorts of combinations of functions we write:
\bea
\Blp \log \blp e^{3\Delta} A_2^2 A_6\brp\Brp' &=& \Blp \log \blp e^{3\Delta} A_1 A_2 A_5\brp\Brp' +\Blp \log \blp  \frac{A_2 A_6}{A_1 A_5}\brp\Brp' \non \\
&=& - e^{-3\Delta}\frac{A_3 A_5}{A_2^2} \\
\Rightarrow \ \ \Blp \log \blp  \frac{A_2 A_6}{A_1 A_5}\brp\Brp' &=&e^{-3\Delta} \Blp -\frac{A_3 A_5}{A_2^2}+\frac{A_3}{A_5} \Brp \non \\
&=&e^{-3\Delta} A_3A_5 \Blp \frac{ 1}{A_5^2}-\frac{1}{A_2^2} \Brp \label{A26on15}
\eea

Now we also need
\bea
p&=&e^{3\Delta}  \blp -A_1^2A_6+A_2^2A_4^2A_6+2A_1A_2 A_4 A_5 A_7  \brp \non \\
&=& e^{3\Delta}\Blp -A_1^2 A_6 +A_2^2 A_6(1-\frac{1}{4}\frac{A_2^2A_6^2}{ A_1^2 A_5^2}) \Brp \label{peq}
\eea
where $p$ is a constant.

\subsubsection{$d\Blp \Re \Xit^{(3)}\Brp$}
The flux is given algebraically by the equation
\bea
\eq{form5}:\ \ e^{-2\Delta}d\Blp e^{2\Delta }\sqrt{\Xi^2-\Xit^2}  \Psi_+\Brp &=& -\Xit  G  \non  \\
\Rightarrow\  -d\Blp  \sqrt{e^{6\Delta}-1}  \Psi_+\Brp &=& G  \non  
\eea
With the previous algebraic results, we have
\bea
\Psi_+&=& \Blp A_1^2 A_5 +\frac{A_2^2 A_6^2}{2A_5}-\frac{A_2^4 A_6^2}{4 A_1^2A_5}+\frac{A_2^6 A_6^4}{8A_1^4A_5^3}\Brp \sig_{123} \non \\
&&  +A_2^2A_5 \Sig_{123} \non \\
&& +\blp-A_1^2 A_5 + \frac{A_2^4 A_6^2}{4 A_1^2 A_5}\brp\sig_{12}\Sig_3  \non \\
&&+\Blp -A_1A_2 A_6 + \frac{ A_2^3 A_6}{2A_1} -\frac{A_2^5 A_6^3}{4 A_1^3 A_5^2}  \Brp (\sig_{23}\Sig_1 + \sig_{31}\Sig_2)  \non \\
&&+ \Blp- A_2^2 A_5 +\frac{A_2^4 A_6^2}{2 A_1^2 A_5} \Brp \sig_3 \Sig_{12} \non \\
&&- \frac{A_2^3 A_6}{2 A_1} \blp \sig_2 \Sig_{31}+ \sig_1 \Sig_{23} \brp \non
\eea
and so the flux is given by
\bea
G_{(4)} &=&-\Blp  \sqrt{e^{6\Delta}-1} \Blp A_1^2 A_5 +\frac{A_2^2 A_6^2}{2A_5}-\frac{A_2^4 A_6^2}{4 A_1^2A_5}+\frac{A_2^6 A_6^4}{8A_1^4A_5^3}\Brp\Brp' dr\w \sig_{123} \non \\
&& -\Blp  \sqrt{e^{6\Delta}-1} A_2^2A_5  \Brp'  dr\w \Sig_{123} \non \\
&& -\Blp  \sqrt{e^{6\Delta}-1}\blp-A_1^2 A_5 + \frac{A_2^4 A_6^2}{4 A_1^2 A_5}\brp  \Brp' dr\w \sig_{12}\Sig_3 \non \\
&& -\Blp  \sqrt{e^{6\Delta}-1}\Blp -A_1A_2 A_6 + \frac{ A_2^3 A_6}{2A_1} -\frac{A_2^5 A_6^3}{4 A_1^3 A_5^2}  \Brp\Brp 'dr\w(\sig_{23}\Sig_1 + \sig_{31}\Sig_2) \non \\
&&-\Blp\sqrt{e^{6\Delta}-1} \Blp- A_2^2 A_5 +\frac{A_2^4 A_6^2}{2 A_1^2 A_5} \Brp\Brp' dr \sig_3 \Sig_{12} \non \\
&&+ \Blp \sqrt{e^{6\Delta}-1} \frac{A_2^3 A_6}{2 A_1} \Brp' dr\w\blp \sig_2 \Sig_{31}+ \sig_1 \Sig_{23} \brp \non\\
&&  -\sqrt{e^{6\Delta}-1}\Blp-A_1^2 A_5 + \frac{A_2^4 A_6^2}{4 A_1^2 A_5}-A_2^2 A_5 -\frac{A_2^4 A_6^2}{2 A_1^2 A_5}  \Brp \sig_{12} \Sig_{12 }\non \\
&&  -\sqrt{e^{6\Delta}-1}\Blp  -A_1A_2 A_6 + \frac{ A_2^3 A_6}{2A_1} -\frac{A_2^5 A_6^3}{4 A_1^3 A_5^2}  - \frac{A_2^3 A_6}{2 A_1}\Brp  \blp \sig_{13} \Sig_{13}+ \sig_{23} \Sig_{2 3}  \brp  \label{G41}
\eea
\subsubsection{$d\Blp \Xi^{(2)}\Brp$}
Now  we seperately calculate the Hodge dual of the four form flux so that we can equate \eq{form3} with \eq{G41}. This will give the last of the equations.
\bea
\eq{form3}\ \ \ \ \ e^{-4\Delta} d\blp  e^{4\Delta} \sqrt{1-e^{-6\Delta}} J \brp &=&-*G_{(4)} \non \\
\Rightarrow G_{(4)} &=&-*e^{-4\Delta} d\blp  e^{4\Delta} \sqrt{1-e^{-6\Delta}} J \brp  \label{HG}.
\eea
We have 
\bea
J &=& -A_1A_2(\sig_1 \Sig_1+\sig_2 \Sig_2) -A_5 A_6 \sig_3 \Sig_3 \non 
\eea
so that
\bea
{\rm RHS}\ \eq{HG} 
&=& *\Blp e^{-4\Delta} \blp   e^{4\Delta} \sqrt{1-e^{-6\Delta}}  A_1A_2 \brp'dr\w  (\sig_1 \Sig_1+\sig_2 \Sig_2) \non \\
&&  + e^{-4\Delta} \blp   e^{4\Delta} \sqrt{1-e^{-6\Delta}}  A_5A_6 \brp'dr\w  \sig_3 \Sig_3 \non \\
&& + \sqrt{1-e^{-6\Delta}}  A_1A_2 d (\sig_1 \Sig_1+\sig_2 \Sig_2) \non \\
&&+ \sqrt{1-e^{-6\Delta}}  A_5A_6 d (\sig_3 \Sig_3)\Brp \non 
\eea

Doing the Hodge dualizing we find
\bea
* dr\w (\sig_1\w \Sig_1+\sig_2\w \Sig_2) &=& \frac{A_5 A_6}{A_3} \Blp \sig_{13} \Sig_{13}+ \sig_{23} \Sig_{23} \Brp \non \\
*dr\w \sig_3 \w \Sig_3 &=&\frac{A_1^2 A_2^2}{A_3 A_5 A_6} \sig_{12}\Sig_{12} \non
\eea
and the more complicated ones are
\bea
* d(\sig_1 \Sig_1+\sig_2 \Sig_2) &=& \Blp -\frac{A_2 A_3 A_6^2}{A_1 A_5^2} + \frac{A_2^3 A_3 A_6^2}{
 2 A_1^3 A_5^2} - \frac{A_2^5 A_3 A_6^4}{2 A_1^5 A_5^4} + \frac{
 A_2^7 A_3 A_6^4}{8 A_1^7 A_5^4} - \frac{
 A_2^9 A_3 A_6^6}{16 A_1^9 A_5^6}
\Brp  dr\sig_{123} \non \\
&&\!\!\!\!\!\!\!\!\!\!\!\!\!\!- \frac{A_2^5 A_3 A_6^2}{2 A_1^5 A_5^2}dr\Sig_{123} \non \\
&&\!\!\!\!\!\!\!\!\!\!\!\!\!\!- \Blp \frac{A_2^3 A_3 A_6^2}{2 A_1^3 A_5^2} + \frac{A_2^7 A_3 A_6^4}{ 8 A_1^7 A_5^4} \Brp dr\sig_{12}\Sig_3 \non \\
&&\!\!\!\!\!\!\!\!\!\!\!\!\!\!+ \Blp \frac{A_3 A_6}{A_5} - \frac{A_2^2 A_3 A_6}{2 A_1^2 A_5}+ \frac{3 A_2^4 A_3 A_6^3}{4 A_1^4 A_5^3}  - \frac{A_2^6 A_3 A_6^3}{4 A_1^6 A_5^3} + \frac{A_2^8 A_3 A_6^5 }{8 A_1^8 A_5^5}  \Brp dr(\sig_{23}\Sig_1 +\sig_{31}\Sig_2)\non \\
&&\!\!\!\!\!\!\!\!\!\!\!\!\!\!+ \Blp -\frac{A_2^3 A_3 A_6^2}{A_1^3 A_5^2} + \frac{A_2^5A_3 A_6^2}{2 A_1^5 A_5^2} - \frac{A_2^7 A_3 A_6^4}{4 A_1^7 A_5^4} \Brp dr\sig_3\Sig_{12} \non \\
&&\!\!\!\!\!\!\!\!\!\!\!\!\!\!+ \Blp \frac{A_2^2 A_3 A_6}{2 A_1^2 A_5} +\frac{A_2^6 A_3 A_6^3}{4 A_1^6 A_5^3} \Brp dr(\sig_1\Sig_{23} +\sig_2 \Sig_{31})\non 
\eea
\bea
* d(\sig_3 \Sig_3) &=& \Blp -\frac{A_1^2 A_3 A_5}{A_2^2 A_6} + \frac{A_3 A_6}{2 A_5} - \frac{
 A_2^2 A_3 A_6}{2 A_1^2 A_5} + \frac{A_2^4 A_3 A_6}{
 4 A_1^4 A_5} \non \\
 && + \frac{A_2^4 A_3 A_6^3}{2 A_1^4 A_5^3}- \frac{
 5 A_2^6 A_3 A_6^3}{16 A_1^6 A_5^3} + \frac{
 3 A_2^8 A_3 A_6^5}{32 A_1^8 A_5^5} \Brp  dr\sig_{123} \non \\
&&+ \Blp -\frac{A_2^2 A_3 A_5}{A_1^2 A_6} + \frac{3 A_2^4 A_3 A_6}{
 4 A_1^4 A_5} \Brp dr\Sig_{123} \non \\
&&+ \Blp \frac{A_1^2 A_3 A_5}{A_2^2 A_6} + \frac{A_2^2 A_3 A_6}{
 2 A_1^2 A_5} - \frac{A_2^4 A_3 A_6}{4 A_1^4 A_5} + \frac{
 3 A_2^6 A_3 A_6^3}{16 A_1^6 A_5^3} \Brp dr\sig_{12}\Sig_3 \non \\
&&+ \Blp \frac{A_2 A_3}{2 A_1} - \frac{A_2^3 A_3}{2 A_1^3} - \frac{
 3 A_2^3 A_3 A_6^2}{4 A_1^3 A_5^2} + \frac{
 5 A_2^5 A_3 A_6^2}{8 A_1^5 A_5^2} - \frac{
 3 A_2^7 A_3 A_6^4}{16 A_1^7 A_5^4}
 \Brp dr(\sig_{23}\Sig_1 +\sig_{31}\Sig_2)\non \\
&&+ \Blp \frac{A_2^2 A_3 A_5}{A_1^2 A_6} + \frac{A_2^2 A_3 A_6}{
 A_1^2 A_5} - \frac{5 A_2^4 A_3 A_6}{4 A_1^4 A_5} + \frac{
 3 A_2^6 A_3 A_6^3}{8 A_1^6 A_5^3} \Brp dr\sig_3\Sig_{12} \non \\
&&+ \Blp  -\frac{A_2 A_3}{2 A_1} + \frac{A_2^3 A_3}{2 A_1^3} - \frac{ 3 A_2^5 A_3 A_6^2}{8 A_1^5 A_5^2}\Brp dr(\sig_1\Sig_{23} +\sig_2 \Sig_{31})\non .
\eea
Putting this all together with \eq{HG} we get another expression for $G^{(4)}$ which we equate with \eq{G41}. 
\subsection{Summary Of Equations}

Here we will summarize all the equations from \eq{Xi3eq1}, \eq{A26on15}, \eq{peq} and from equating   \eq{HG} with \eq{G41}:

\bea
p&=& e^{3\Delta}\Blp -A_1^2 A_6 +A_2^2 A_6(1-\frac{1}{4}\frac{A_2^2A_6^2}{ A_1^2 A_5^2}) \Brp \label{feq1} \\
\Blp \log \blp  \frac{A_2 A_6}{A_1 A_5}\brp\Brp' &=&e^{-3\Delta} A_3A_5 \Blp \frac{ 1}{A_5^2}-\frac{1}{A_2^2} \Brp \label{feq2}\\
\Blp \log \blp e^{3\Delta} A_2^2 A_6\brp\Brp'&=& - e^{-3\Delta}\frac{A_3 A_5}{A_2^2} \label{feq3} \\
\Blp \log \blp e^{3\Delta} A_1 A_2 A_5\brp\Brp'&=& - e^{-3\Delta}\frac{A_3}{A_5}  \label{feq4} \\
\Blp \log \Blp \sqrt{e^{6\Delta}-1} A_2^2 A_5 \Brp\Brp'&=& -e^{-3\Delta} \Blp -\frac{A_3 A_5}{A_1^2} - \frac{A_2^4 A_3 A_6^2}{2 A_1^4 A_5^3} + \frac{3 A_2^2 A_3 A_6^2}{4 A_1^4 A_5}  \Brp \label{feq5}\\
\Blp \log \Blp \sqrt{e^{6\Delta}-1} A_1^2 A_5 \Brp\Brp'&=& e^{-3\Delta} \Blp
\frac{A_3 A_5}{A_2^2} \Brp \label{feq6} \\
 \blp   e^{4\Delta} \sqrt{1-e^{-6\Delta}}  A_1A_2 \brp'  &=& -e^{4\Delta}\sqrt{e^{6\Delta}-1}\frac{A_2 A_3}{A_5 }\Blp  -A_1    -\frac{A_2^4 A_6^2}{4 A_1^3 A_5^2} \Brp   \label{extra1}
\eea

\subsection{Solving the Equations}
First we redefine thefive modes $(A_1,A_2,A_5,A_6,\Delta)$ in terms of $(f_!,f_2,g_1,g_2,g_3)$:
\bea
f_1=e^{3\Delta} A_2^2 A_6,&&
f_2=\sqrt{e^{6\Delta}-1} A_1^2 A_5,\non \\
g_1= \frac{A_2^2}{A_1^2},&&g_2= \frac{A_2^2}{A_5^2},\ \ \ \ g_3= \frac{A_2 A_6}{A_1 A_5} .\non 
\eea
In fact $g_3=A_4$ which we eliminated  earlier.

To solve the equations, we first use the infinite wisdom of hindsight to choose a radial co-ordinate such that
\be
A_3= -e^{3\Delta} \frac{A_2^2}{A_5}\frac{2}{r},
\ee
this allows us to integrate \eq{feq3} and \eq{feq6} and we find
\bea
f_1 &=&c_1 r^2, \non \\
f_2 &=&\frac{c_2}{r^2}. \non 
\eea
Then we use \eq{feq1} to solve algebraically for $g_1$ (with $p=8r_0^2$):
\be
\frac{1}{g_1}= 1-\frac{1}{4}g_3^2  +\frac{8r_0^2}{c_1 r^2} \label{geq1} 
\ee
and we are left with
\bea
\eq{feq2}:\ \ \ \ \ \ \ \ \ \ \Blp \log g_3 \Brp' &=&\frac{2}{r} \Blp 1-g_2  \Brp  \label{geq2}\\
\eq{feq5}:\ \ \ \ \ \ \Blp \log (f_2 g_1)\Brp'&=& - \frac{2g_1}{r}\Blp 1+\half g_3^2 (g_2 -3/2) \Brp
\eea
We can work out that 
\bea
g_1&=& \frac{4r^2}{4(r^2+ 8r_0^2/c_1)-r^2 g_3^2} \non \\
f_2g_1&=& \frac{4 c_2}{4(r^2+ 8r_0^2/c_1)-r^2 g_3^2} \\
g_2&=&1- \frac{(\log\, g_3)'}{r}  \non 
\eea
The other equations are now satisfied as well

At this point our solution is
\bea
e^{3\Delta} A_2^2 A_6 &=&c_1 r^2 \non \\
\sqrt{e^{6\Delta}-1} A_1^2 A_5&=&\frac{c_2}{r^2}\non \\
  \frac{A_2^2}{A_1^2}&=&  \frac{4r^2}{4(r^2+ 8r_0^2/c_1)-r^2 g_3^2} \non \\
  \frac{A_2^2}{A_5^2},&=&1- \frac{(\log\, g_3)'}{r}   \\
 g_3&=& \frac{A_2 A_6}{A_1 A_5} \non \\
 A_4&=&\half g_3 \non \\
 A_7 &=& 1-\half g_3^2. \non 
\eea
We can then eliminate $c_1$ by rescaling the radial co-ordinate $r$ and we will define $c=c_1c_2$.

\section{Solutions of Massive IIA}
Here we will make an ansatz for massive IIA supergravity with the symmetries \eq{sym1} and proceed to solve up to the point where we find that the mass term must vanish. There is no need to proceed further since we have already solved the most general solution with these symmetries in $d=11$ supergravity. 

We will use the supersymmetry conditions of massive IIA as written down in \cite{Grana:2005sn},
the advantage of using differential forms rather than the fermionic variations is essentially
that a lot of work with the Clifford algebra has algebra has already been performed.
The most general spinor ansatz for reducing type II supergrvaity to four dimensions (preserving four supercharges) involves two $SO(6)$ spinors of positive chirality 
\be
\eta_+,\ \ \ \chi_+=z\cdot \eta_-
\ee
where $z=z_mdx^m$ is a one form.
The $SO(1,9)$ spinors are decomposed as
\bea
\eps_+^1&=& \zeta_+ \otimes \eta_+^1 + c.c \non \\
\eps_-^2&=& \zeta_+ \otimes \eta_-^2 + c.c \non 
\eea
It was shown in \cite{Halmagyi:2007ft} that the most general spinor ansatz can be written as
\bea
\begin{array}{rcl}
\eta_+^1&=& c_\phi  \eta_+ + s_{\phi} \chi_+, \\
\eta_+^2&=&e^{i\tha}( c_\phi  \eta_+ - s_{\phi} \chi_+) . 
\end{array} \label{eta12}
\eea
One can then construct the spinor bilinears as particular sums of even and odd differential forms
\bea
\Phi_+ &=& \eta^1_+ \otimes \eta_+^{2 \dagger}\non \\
&=& \frac{1}{8}\Blp a \xbar e^{-ij} + b \ybar e^{ij} -i(a \ybar\, \om + \xbar b\, \ombar)  \Brp\w  e^{z \zbar/2}, \\
\Phi_-&=&\eta^1_+ \otimes \eta_-^{2 \dagger} \non \\
&=& \frac{1}{8} \Blp i(b y \ombar -a x \om) + (bx e^{ij} - ay e^{-ij}) \Brp\w z.
\eea
The string frame supersymmetry conditions can be written in a compact form but of course this must be expanded into components where it will be as cumbersome as usual:
\bea
e^{-2A+\vphi} d_H \blp e^{2A-\vphi} \Phi_+ \brp &=&0\label{EqPhiPl} \\
e^{-2A+\vphi} d_H \blp e^{2A-\vphi} \Phi_- \brp &=& dA \w \ol{\Phi}_- + \frac{i}{16} e^{A+\vphi} \lambda(*F). \label{EqPhiMin}
\eea
where $d_H=d-H\w$ and in IIA we have 
\bea
F&=& F_0 + F_2 +F_4+F_6 \non \\
\lambda(*F)&=& -*F_0+*F_2 - *F_4 + * F_6. \non
\eea
Key to checking the various conventions in the literature regarding $F$ and $d_H$ is the Bianchi identity
\be
d_H F=0 \ \ \Rightarrow \ \ d F_{n}= H\w F_{n-2}.\non 
\ee

\subsection{Massive IIA and $SU(3)$ Structures}

The spinors \eq{eta12} define the most general $SU(2)$ structure but it is worth considering a few sub-cases.  When $\phi=0$ in \eq{eta12} we have the most general $SU(3)$ structure and here it is easy to see that there are no Minkowski solutions with $F_0 \neq 0$.

In this case the spinor bilinears are given by
\bea
\Phi_+&=& \frac{e^{i\tha}}{8}e^{-iJ} \\
\Phi_- &=& \frac{e^{-i\tha}}{8} \Om .
\eea
The one form part of \eq{EqPhiPl} gives 
\be
d(e^{i\tha} e^{3A-\vphi})=0
\ee
from which we discover that $\tha$ is a constant and 
\be
3A=\vphi.
\ee
The three form part of the same equation gives
\be
dJ=0,\ \ \ H=0.
\ee
 Then from \eq{EqPhiMin} we see that $H=0\Rightarrow F_0=0$. It would be interesting to have a general argument that $SU(2)$ structure backgrounds with and four dimensional Minkowski factor must have $F_0=0$ but we have not found such an argument. Nonethless, below we find that this is true for the particular backgrounds we study here.
	
\subsection{The Ansatz}
Our ansatz for the metric is
\be
ds_{10}^2=e^{2A} ds_4^2 + ds_{M_6}^2\non
\ee
where the frames on $M_6$ are given by
\bea
E_1 &=&c^{1/2} _{2 \phi}A_1 (e_1 + A_2 \eps_1) \non \\
E_2&=& c^{1/2} _{2 \phi}A_1 (e_2 - A_2 \eps_2)\non \\
E_3 &=& c^{1/2} _{2 \phi} A_3  (\eps_1 + A_4 e_1)\non \\
E_4&=& c^{1/2} _{2 \phi}A_3 (\eps_2 - A_4 e_2)\non \\
E_5&=& A_5^{-1} dr\non \\
E_6 &=& (r/3) A_5  (e_3 + \eps_3)\non .
\eea
By our choice of frames for $E_5$ and $E_6$ we have defined our radial co-ordinate
to co-incide with that of the conventional  choice for the Ricci flat metric on the conifold.
Then the fluxes (which automically satisfy their Bianchi identities) are given  by
\bea
F_0&=&m \non \\
F_2&=&N e_1\w e_2 + d\Blp f_1(e_3+\eps_3)\Brp +mB \non \\
F_4&=& d\Blp \blp f_2e_1\w e_2+ 
  f_3 \eps_1\w \eps_2+  f_4 (e_1\w \eps_1 + e_2, \eps_2)+ 
  f_5 ( e_1\w\eps_2- e_2\w \eps_1)\brp \w(e_3 + \eps_3)\Brp \non \\
  &&+ B\w F_2 - m B\w B. \non 
\eea
The dilaton $e^\phi$ is arbitrary and we will hold off from making an ansatz for the $H_3$ flux since we will immediately see that it is given algebraically in terms of the metric. Note that $F_4$ has no components along the worldvolume of Minkowski space since such a field strength would automatically give rise to a non-zero cosmological constant.

\subsection{The $\Phi_+$ Equations}
The $\Phi_+$ equations are fairly straightforward. From the one form we get
\bea
d(e^{3A-\vphi} c_{2\phi})&=&0 \non  \\
\Rightarrow \ \ e^{\vphi}&=&\frac{e^{3A}}{ c_{2\phi}} \label{phplus1}
\eea
The three form gives
\be
d J = 0  \label{phplus2}
\ee
where
\be
J=\frac{j}{c_{2\phi}} + \frac{i}{2} z\zbar
\ee
and
\be
H=t_{2\phi} \Im \om  \label{phplus3}.
\ee
The constraints from \eq{phplus2} determine the symplectic form $J$ to be in fact invariant
\bea
A_4&=&\frac{A_1^2A_2}{A_3^2} \non \\
A_3&=& \sqrt{6} A_1^2 A_2(-a^2 - r^2 + 6 A_1^2)^{-1/2} \non \\
A_2 &=& \frac{r (-a^2 - r^2 + 6 A_1^2)^{1/2} }{\sqrt{6} \sqrt{a^2 + r^2}A_1}
\eea
 and thus
 \be
 J=\frac{( a^2 +r^2)}{6} e_1 \w e_2 +  \frac{r^2}{6} \eps_1 \w \eps_2+\frac{2}{3} r dr\w (e_3 +\eps_3).
 \ee

\subsection{The $\Phi_-$ Equations}
This equation for $\Phi_-$ breaks into real and imaginary parts
\bea
e^{\vphi-A} d_H (e^{A-\vphi} \Re \Phi_-) &=&0 ,\label{pmin1} \\
e^{-3A+\vphi}d_H(e^{3A-\vphi} \Im \Phi_-)  &=&  \frac{e^{A+\vphi}}{8} \lambda(*F).
\eea
\subsubsection{$\Re \Phi_-$}
The equation for $\Re \Phi_-$ is fairly straightforward, the two form component determines the phase of $z$, so that 
\be
\Im z\sim dr.
\ee
The four form component is gives one non-trivial equation
\be
\Blp e^{-A} r^2 \sqrt{a^2+r^2}A_5\Brp'= \frac{3 e^{-A} r \sqrt{a^2+r^2} \cosh(2B_1)}{c_{2\phi}A_5}
\ee
\subsubsection{$\Im \Phi_-$}
Since the equations $\Im \Phi_-$ depend on the Ramond flux, they are significantly more complicated than the others. From the two form part we find several relations, firstly we find some constraints on the $F_4$ flux:
\be
f_4=0,\ \ f_5=0.
\ee
Then we have several equations on the remaining functions
\bea
f'_3&=& \frac{e^{-3 A} r^4 A_5 t_{2 \phi}}{9 (a^2 + r^2)}  \non \\
 (a^2 + r^2) s_{2 \phi} 
f'_1+ 6 \sinh(2 B_1) f'_2&=&-\frac{2 r^2 A_5 \sinh(2 B_1) t_{2 \phi}}{3 e^{3 A}} \\ 
e^{-4A} \Blp e^{A} r A_5 t_{2 \phi} \Brp'&=&-\frac{36 (f_2 + f_3) }{c_{2\phi}^2} + 
      6 (N r^2 - a^2 f_1) \sinh(2B_1) \frac{s_{2 \phi}}{c_{2\phi}^2}+ m r t_{2 \phi}^2 \non 
\eea

From the four form we get three independent equations
\bea
f'_1&=& \frac{ 2 a^2 e^{-3 A} A_5  \sinh(2 B_1)}{ 3 (a^2 + r^2)c_{2 \phi}} \\
  g'_2 + e^A r A_5 \tan(2\phi) g'_3 &=& = \frac{6N e^{4A}r^3 \cosh^2 (2B_1)}{a^2 +r^2} + m e^{4A}r^3 \sin (2\phi) \sinh(2B_1)\non \\
  &&+ \frac{6 e^{4A}r f_1 (r^2-a^2 \sinh^2(2B_1))}{a^2 +r^2}\\
\cos^2(2\phi) g_5 g'_4 + g_4 g'_5  &=& \frac{3e^A r^2 \cos(2\phi)}{A_5}  + \frac{a^2 A_5 \cosh(2B_1)}{(a^2 + r^2)} - \half e^{3A} \cos(4\phi) \cosh(2B_1) m r^2 \non \\
&&+ \frac{e^{3A} \cos(2\phi) 6(Nr^2-a^2 f_1)\sinh(4B_1)}{2(a^2 +r^2)}   
\eea
where
\bea
g_2&=& e^A r^3 A_5 \cos(2\phi) \sinh(2B_1)  \non \\
g_3&=& r^2 \sin(2\phi) \sinh(2B_1) \non \\
g_4&=& e^A r A_5 \non \\
g_5 &=& r^2 \cosh(2B_1) \non.
\eea
From the six form, we obviously just get one equation and it is
\be
\Blp e^A r A_5 \cot(2\phi)\Brp'=-e^{4 A} m r \cot^2(2 \phi) + \frac{ 6 e^A \cosh(2 B_1) }{s_{2 \phi}A_5} .
\ee

\subsection{Equation Analysis}

One finds from analyzing these equations that
\bea
f_1&=& -\frac{Nr^2}{2(a^2+r^2)}, \non \\
A_5&=& -\frac{3e^3A Nr \cos(2\phi)}{2(a^2 +r^2)\sinh(2B_1)} .\non
\eea
With some amount of work we used the remaining equations to find two different expressions for $B'_1$ which can only be equal if $m=0$. At this point it is not worthwhile to continue since any solutions found would be already in our analysis from $d=11$ supergravity.
\end{appendix}
\bibliographystyle{/Users/Halmagyi/utphys} 
\bibliography{/Users/Halmagyi/myrefs}

\providecommand{\href}[2]{#2}\begingroup\raggedright\begin{thebibliography}{10}

\bibitem{Strominger:1995cz}
A.~Strominger, ``{Massless black holes and conifolds in string theory},'' {\em
  Nucl. Phys.} {\bf B451} (1995) 96--108,
\href{http://arXiv.org/abs/hep-th/9504090}{{\tt hep-th/9504090}}.

\bibitem{Candelas:1989js}
P.~Candelas and X.~C. de~la Ossa, ``Comments on conifolds,'' {\em Nucl. Phys.}
  {\bf B342} (1990)
246--268.

\bibitem{Klebanov:1998hh}
I.~R. Klebanov and E.~Witten, ``Superconformal field theory on threebranes at a
  calabi-yau singularity,'' {\em Nucl. Phys.} {\bf B536} (1998) 199--218,
\href{http://arXiv.org/abs/hep-th/9807080}{{\tt hep-th/9807080}}.

\bibitem{Klebanov:2000hb}
I.~R. Klebanov and M.~J. Strassler, ``Supergravity and a confining gauge
  theory: Duality cascades and chisb-resolution of naked singularities,'' {\em
  JHEP} {\bf 08} (2000) 052,
\href{http://arXiv.org/abs/hep-th/0007191}{{\tt hep-th/0007191}}.

\bibitem{Maldacena:2000mw}
J.~M. Maldacena and C.~Nunez, ``Supergravity description of field theories on
  curved manifolds and a no go theorem,'' {\em Int. J. Mod. Phys.} {\bf A16}
  (2001) 822--855,
\href{http://arXiv.org/abs/hep-th/0007018}{{\tt hep-th/0007018}}.

\bibitem{Gopakumar:1998ki}
R.~Gopakumar and C.~Vafa, ``On the gauge theory/geometry correspondence,'' {\em
  Adv. Theor. Math. Phys.} {\bf 3} (1999) 1415--1443,
\href{http://arXiv.org/abs/hep-th/9811131}{{\tt hep-th/9811131}}.

\bibitem{Dijkgraaf:2002fc}
R.~Dijkgraaf and C.~Vafa, ``Matrix models, topological strings, and
  supersymmetric gauge theories,'' {\em Nucl. Phys.} {\bf B644} (2002) 3--20,
\href{http://arXiv.org/abs/hep-th/0206255}{{\tt hep-th/0206255}}.

\bibitem{Vafa:2000wi}
C.~Vafa, ``Superstrings and topological strings at large n,'' {\em J. Math.
  Phys.} {\bf 42} (2001) 2798--2817,
\href{http://arXiv.org/abs/hep-th/0008142}{{\tt hep-th/0008142}}.

\bibitem{Ooguri:1996me}
H.~Ooguri and C.~Vafa, ``Summing up d-instantons,'' {\em Phys. Rev. Lett.} {\bf
  77} (1996) 3296--3298,
\href{http://arXiv.org/abs/hep-th/9608079}{{\tt hep-th/9608079}}.

\bibitem{RoblesLlana:2007ae}
D.~Robles-Llana, F.~Saueressig, U.~Theis, and S.~Vandoren, ``{Membrane
  instantons from mirror symmetry},''
\href{http://arXiv.org/abs/0707.0838}{{\tt 0707.0838}}.

\bibitem{Gubser:2000vg}
S.~S. Gubser, ``Supersymmetry and f-theory realization of the deformed conifold
  with three-form flux,''
\href{http://arXiv.org/abs/hep-th/0010010}{{\tt hep-th/0010010}}.

\bibitem{Grana:2001xn}
M.~Grana and J.~Polchinski, ``Gauge / gravity duals with holomorphic dilaton,''
  {\em Phys. Rev.} {\bf D65} (2002) 126005,
\href{http://arXiv.org/abs/hep-th/0106014}{{\tt hep-th/0106014}}.

\bibitem{Giddings:2001yu}
S.~B. Giddings, S.~Kachru, and J.~Polchinski, ``Hierarchies from fluxes in
  string compactifications,'' {\em Phys. Rev.} {\bf D66} (2002) 106006,
\href{http://arXiv.org/abs/hep-th/0105097}{{\tt hep-th/0105097}}.

\bibitem{Gubser:2004qj}
S.~S. Gubser, C.~P. Herzog, and I.~R. Klebanov, ``Symmetry breaking and axionic
  strings in the warped deformed conifold,'' {\em JHEP} {\bf 09} (2004) 036,
\href{http://arXiv.org/abs/hep-th/0405282}{{\tt hep-th/0405282}}.

\bibitem{Butti:2004pk}
A.~Butti, M.~Grana, R.~Minasian, M.~Petrini, and A.~Zaffaroni, ``The baryonic
  branch of klebanov-strassler solution: A supersymmetric family of su(3)
  structure backgrounds,'' {\em JHEP} {\bf 03} (2005) 069,
\href{http://arXiv.org/abs/hep-th/0412187}{{\tt hep-th/0412187}}.

\bibitem{Casero:2006pt}
R.~Casero, C.~Nunez, and A.~Paredes, ``{Towards the string dual of N = 1
  SQCD-like theories},'' {\em Phys. Rev.} {\bf D73} (2006) 086005,
\href{http://arXiv.org/abs/hep-th/0602027}{{\tt hep-th/0602027}}.

\bibitem{Strominger:1986uh}
A.~Strominger, ``Superstrings with torsion,'' {\em Nucl. Phys.} {\bf B274}
  (1986)
253.

\bibitem{Hull:1986kz}
C.~M. Hull, ``{Compactifications Of The Heterotic Superstring},'' {\em Phys.
  Lett.} {\bf B178} (1986)
357.

\bibitem{Maldacena:2009mw}
J.~Maldacena and D.~Martelli, ``{The unwarped, resolved, deformed conifold:
  fivebranes and the baryonic branch of the Klebanov-Strassler theory},''
\href{http://arXiv.org/abs/0906.0591}{{\tt 0906.0591}}.

\bibitem{Acharya:2000gb}
B.~S. Acharya, ``On realising n = 1 super yang-mills in m theory,''
\href{http://arXiv.org/abs/hep-th/0011089}{{\tt hep-th/0011089}}.

\bibitem{Atiyah:2000zz}
M.~Atiyah, J.~M. Maldacena, and C.~Vafa, ``An m-theory flop as a large n
  duality,'' {\em J. Math. Phys.} {\bf 42} (2001) 3209--3220,
\href{http://arXiv.org/abs/hep-th/0011256}{{\tt hep-th/0011256}}.

\bibitem{Brandhuber:2001kq}
A.~Brandhuber, ``{G(2) holonomy spaces from invariant three-forms},'' {\em
  Nucl. Phys.} {\bf B629} (2002) 393--416,
\href{http://arXiv.org/abs/hep-th/0112113}{{\tt hep-th/0112113}}.

\bibitem{Klebanov:2000nc}
I.~R. Klebanov and A.~A. Tseytlin, ``{Gravity duals of supersymmetric SU(N) x
  SU(N+M) gauge theories},'' {\em Nucl. Phys.} {\bf B578} (2000) 123--138,
\href{http://arXiv.org/abs/hep-th/0002159}{{\tt hep-th/0002159}}.

\bibitem{Linch:2006ig}
W.~D. Linch~III and B.~C. Vallilo, ``Hybrid formalism, supersymmetry reduction,
  and ramond- ramond fluxes,''
\href{http://arXiv.org/abs/hep-th/0607122}{{\tt hep-th/0607122}}.

\bibitem{Berkovits:1996bf}
N.~Berkovits, ``A new description of the superstring,''
\href{http://arXiv.org/abs/hep-th/9604123}{{\tt hep-th/9604123}}.

\bibitem{Gurrieri:2002wz}
S.~Gurrieri, J.~Louis, A.~Micu, and D.~Waldram, ``Mirror symmetry in
  generalized calabi-yau compactifications,'' {\em Nucl. Phys.} {\bf B654}
  (2003) 61--113,
\href{http://arXiv.org/abs/hep-th/0211102}{{\tt hep-th/0211102}}.

\bibitem{Grana:2006hr}
M.~Grana, J.~Louis, and D.~Waldram, ``{SU(3) x SU(3) compactification and
  mirror duals of magnetic fluxes},'' {\em JHEP} {\bf 04} (2007) 101,
\href{http://arXiv.org/abs/hep-th/0612237}{{\tt hep-th/0612237}}.

\bibitem{Strominger:1996it}
A.~Strominger, S.-T. Yau, and E.~Zaslow, ``Mirror symmetry is t-duality,'' {\em
  Nucl. Phys.} {\bf B479} (1996) 243--259,
\href{http://arXiv.org/abs/hep-th/9606040}{{\tt hep-th/9606040}}.

\bibitem{Fidanza:2003zi}
S.~Fidanza, R.~Minasian, and A.~Tomasiello, ``Mirror symmetric su(3)-structure
  manifolds with ns fluxes,'' {\em Commun. Math. Phys.} {\bf 254} (2005)
  401--423,
\href{http://arXiv.org/abs/hep-th/0311122}{{\tt hep-th/0311122}}.

\bibitem{Tomasiello:2005bp}
A.~Tomasiello, ``Topological mirror symmetry with fluxes,'' {\em JHEP} {\bf 06}
  (2005) 067,
\href{http://arXiv.org/abs/hep-th/0502148}{{\tt hep-th/0502148}}.

\bibitem{Becker:2004qh}
M.~Becker, K.~Dasgupta, A.~Knauf, and R.~Tatar, ``{Geometric transitions, flops
  and non-Kaehler manifolds. I},'' {\em Nucl. Phys.} {\bf B702} (2004)
  207--268,
\href{http://arXiv.org/abs/hep-th/0403288}{{\tt hep-th/0403288}}.

\bibitem{Hellerman:2002ax}
S.~Hellerman, J.~McGreevy, and B.~Williams, ``Geometric constructions of
  nongeometric string theories,'' {\em JHEP} {\bf 01} (2004) 024,
\href{http://arXiv.org/abs/hep-th/0208174}{{\tt hep-th/0208174}}.

\bibitem{Bouwknegt:2004ap}
P.~Bouwknegt, K.~Hannabuss, and V.~Mathai, ``Nonassociative tori and
  applications to t-duality,'' {\em Commun. Math. Phys.} {\bf 264} (2006)
  41--69,
\href{http://arXiv.org/abs/hep-th/0412092}{{\tt hep-th/0412092}}.

\bibitem{Dabholkar:2005ve}
A.~Dabholkar and C.~Hull, ``Generalised t-duality and non-geometric
  backgrounds,'' {\em JHEP} {\bf 05} (2006) 009,
\href{http://arXiv.org/abs/hep-th/0512005}{{\tt hep-th/0512005}}.

\bibitem{Chamseddine:1997nm}
A.~H. Chamseddine and M.~S. Volkov, ``{Non-Abelian BPS monopoles in N = 4
  gauged supergravity},'' {\em Phys. Rev. Lett.} {\bf 79} (1997) 3343--3346,
\href{http://arXiv.org/abs/hep-th/9707176}{{\tt hep-th/9707176}}.

\bibitem{Maldacena:2000yy}
J.~M. Maldacena and C.~Nunez, ``Towards the large n limit of pure n = 1 super
  yang mills,'' {\em Phys. Rev. Lett.} {\bf 86} (2001) 588--591,
\href{http://arXiv.org/abs/hep-th/0008001}{{\tt hep-th/0008001}}.

\bibitem{Gauntlett:2003cy}
J.~P. Gauntlett, D.~Martelli, and D.~Waldram, ``Superstrings with intrinsic
  torsion,'' {\em Phys. Rev.} {\bf D69} (2004) 086002,
\href{http://arXiv.org/abs/hep-th/0302158}{{\tt hep-th/0302158}}.

\bibitem{Dymarsky:2005xt}
A.~Dymarsky, I.~R. Klebanov, and N.~Seiberg, ``{On the moduli space of the
  cascading SU(M+p) x SU(p) gauge theory},'' {\em JHEP} {\bf 01} (2006) 155,
\href{http://arXiv.org/abs/hep-th/0511254}{{\tt hep-th/0511254}}.

\bibitem{PandoZayas:2000sq}
L.~A. Pando~Zayas and A.~A. Tseytlin, ``{3-branes on resolved conifold},'' {\em
  JHEP} {\bf 11} (2000) 028,
\href{http://arXiv.org/abs/hep-th/0010088}{{\tt hep-th/0010088}}.

\bibitem{Bena:2009xk}
I.~Bena, M.~Grana, and N.~Halmagyi, ``{On the Existence of Meta-stable Vacua in
  Klebanov- Strassler},''
\href{http://arXiv.org/abs/0912.3519}{{\tt 0912.3519}}.

\bibitem{Candelas:1984yd}
P.~Candelas and D.~J. Raine, ``Spontaneous compactification and supersymmetry
  in d = 11 supergravity,'' {\em Nucl. Phys.} {\bf B248} (1984)
415.

\bibitem{bryant}
R.~L. Bryant and S.~Salamon, ``{On the Construction of some Complete Metrics
  with Exceptional Holonomy},'' {\em Duke Math. J.} {\bf 58} (1989) 829.

\bibitem{Gibbons:1989er}
G.~W. Gibbons, D.~N. Page, and C.~N. Pope, ``{Einstein Metrics on S**3 R**3 and
  R**4 Bundles},'' {\em Commun. Math. Phys.} {\bf 127} (1990)
529.

\bibitem{Hitchin:2004ut}
N.~Hitchin, ``Generalized calabi-yau manifolds,'' {\em Quart. J. Math. Oxford
  Ser.} {\bf 54} (2003) 281--308,
\href{http://arXiv.org/abs/math.dg/0209099}{{\tt math.dg/0209099}}.

\bibitem{Gualtieri:2003dx}
M.~Gualtieri, ``Generalized complex geometry,''
\href{http://arXiv.org/abs/math.dg/0401221}{{\tt math.dg/0401221}}.

\bibitem{Hull:2004in}
C.~M. Hull, ``A geometry for non-geometric string backgrounds,'' {\em JHEP}
  {\bf 10} (2005) 065,
\href{http://arXiv.org/abs/hep-th/0406102}{{\tt hep-th/0406102}}.

\bibitem{Shelton:2005cf}
J.~Shelton, W.~Taylor, and B.~Wecht, ``Nongeometric flux compactifications,''
  {\em JHEP} {\bf 10} (2005) 085,
\href{http://arXiv.org/abs/hep-th/0508133}{{\tt hep-th/0508133}}.

\bibitem{Halmagyi:2008dr}
N.~Halmagyi, ``{Non-geometric String Backgrounds and Worldsheet Algebras},''
  {\em JHEP} {\bf 07} (2008) 137,
\href{http://arXiv.org/abs/0805.4571}{{\tt 0805.4571}}.

\bibitem{Halmagyi:2009te}
N.~Halmagyi, ``{Non-geometric Backgrounds and the First Order String Sigma
  Model},''
\href{http://arXiv.org/abs/0906.2891}{{\tt 0906.2891}}.

\bibitem{Koerber:2007xk}
P.~Koerber and L.~Martucci, ``{From ten to four and back again: how to
  generalize the geometry},'' {\em JHEP} {\bf 08} (2007) 059,
\href{http://arXiv.org/abs/0707.1038}{{\tt 0707.1038}}.

\bibitem{Gauntlett:2002fz}
J.~P. Gauntlett and S.~Pakis, ``The geometry of d = 11 killing spinors,'' {\em
  JHEP} {\bf 04} (2003) 039,
\href{http://arXiv.org/abs/hep-th/0212008}{{\tt hep-th/0212008}}.

\bibitem{Pilch:2003jg}
K.~Pilch and N.~P. Warner, ``Generalizing the n = 2 supersymmetric rg flow
  solution of iib supergravity,'' {\em Nucl. Phys.} {\bf B675} (2003) 99--121,
\href{http://arXiv.org/abs/hep-th/0306098}{{\tt hep-th/0306098}}.

\bibitem{Bena:2004jw}
I.~Bena and N.~P. Warner, ``A harmonic family of dielectric flow solutions with
  maximal supersymmetry,'' {\em JHEP} {\bf 12} (2004) 021,
\href{http://arXiv.org/abs/hep-th/0406145}{{\tt hep-th/0406145}}.

\bibitem{Halmagyi:2005pn}
N.~Halmagyi, K.~Pilch, C.~Romelsberger, and N.~P. Warner, ``Holographic duals
  of a family of n = 1 fixed points,''
\href{http://arXiv.org/abs/hep-th/0506206}{{\tt hep-th/0506206}}.

\bibitem{Fu:2006vj}
J.-X. Fu and S.-T. Yau, ``The theory of superstring with flux on non-kaehler
  manifolds and the complex monge-ampere equation,''
\href{http://arXiv.org/abs/hep-th/0604063}{{\tt hep-th/0604063}}.

\bibitem{Grana:2005sn}
M.~Grana, R.~Minasian, M.~Petrini, and A.~Tomasiello, ``Generalized structures
  of n = 1 vacua,'' {\em JHEP} {\bf 11} (2005) 020,
\href{http://arXiv.org/abs/hep-th/0505212}{{\tt hep-th/0505212}}.

\bibitem{Tomasiello:2007zq}
A.~Tomasiello, ``Reformulating supersymmetry with a generalized dolbeault
  operator,''
\href{http://arXiv.org/abs/arXiv:0704.2613 [hep-th]}{{\tt arXiv:0704.2613
  [hep-th]}}.

\bibitem{Lukas:2004ip}
A.~Lukas and P.~M. Saffin, ``{M-theory compactification, fluxes and AdS(4)},''
  {\em Phys. Rev.} {\bf D71} (2005) 046005,
\href{http://arXiv.org/abs/hep-th/0403235}{{\tt hep-th/0403235}}.

\bibitem{Kaste:2003zd}
P.~Kaste, R.~Minasian, and A.~Tomasiello, ``Supersymmetric m-theory
  compactifications with fluxes on seven-manifolds and g-structures,'' {\em
  JHEP} {\bf 07} (2003) 004,
\href{http://arXiv.org/abs/hep-th/0303127}{{\tt hep-th/0303127}}.

\bibitem{Dall'Agata:2003ir}
G.~Dall'Agata and N.~Prezas, ``N = 1 geometries for m-theory and type iia
  strings with fluxes,'' {\em Phys. Rev.} {\bf D69} (2004) 066004,
\href{http://arXiv.org/abs/hep-th/0311146}{{\tt hep-th/0311146}}.

\bibitem{Gauntlett:2004zh}
J.~P. Gauntlett, D.~Martelli, J.~Sparks, and D.~Waldram, ``Supersymmetric
  ads(5) solutions of m-theory,'' {\em Class. Quant. Grav.} {\bf 21} (2004)
  4335--4366,
\href{http://arXiv.org/abs/hep-th/0402153}{{\tt hep-th/0402153}}.

\bibitem{Gran:2001yh}
U.~Gran, ``{GAMMA: A Mathematica package for performing Gamma-matrix algebra
  and Fierz transformations in arbitrary dimensions},''
\href{http://arXiv.org/abs/hep-th/0105086}{{\tt hep-th/0105086}}.

\bibitem{Halmagyi:2007ft}
N.~Halmagyi and A.~Tomasiello, ``Generalized kaehler potentials from
  supergravity,''
\href{http://arXiv.org/abs/arXiv:0708.1032 [hep-th]}{{\tt arXiv:0708.1032
  [hep-th]}}.

\end{thebibliography}\endgroup

\end{document}